\documentclass{article}

\usepackage{PRIMEarxiv}

\usepackage{isomath}
\usepackage{amsthm}
\usepackage{amsmath}
\usepackage{amsfonts}

\usepackage{pifont}
\usepackage{amssymb}
\usepackage{verbatim}
\usepackage{upgreek}

\usepackage{color}
\usepackage{epsfig}
\usepackage{eucal}
\usepackage{subfigure}
\usepackage{cuted}

\usepackage[utf8]{inputenc} 
\usepackage[T1]{fontenc}    
\usepackage{hyperref}       
\usepackage{url}            
\usepackage{booktabs}       
\usepackage{amsfonts}       
\usepackage{nicefrac}       
\usepackage{microtype}      
\usepackage{lipsum}
\usepackage{fancyhdr}       
\usepackage{graphicx}       
\graphicspath{{media/}}     

\title{Synthetic Aperture Radar Doppler Tomography Reveals Details of Undiscovered High-Resolution Internal Structure of the Great Pyramid of Giza
\thanks{\textit{\underline{Citation}}: 
\textbf{Authors. Title. Pages.... DOI:000000/11111.}} 
}

\author{
  Filippo Biondi $^{1,\dagger,\ddagger}$ \\
  Department of Electronic and Electrical Engineering,  \\
  University of Strathclyde \\
  Glasgow (U.K.)\\
  \texttt{filippo.biondi@strath.ac.uk} \\
   \And
  Corrado Malanga \\
   Department of Chemistry and Industrial Chemistry,  \\
  University of Pisa Italy,  \\
  Via Giuseppe Moruzzi, 13, 56100 Pisa,\\
  \texttt{corrado.malanga@unipi.it} \\
}

\begin{document}
\maketitle

\begin{abstract}
A problem with synthetic aperture radar (SAR) is that, due to the poor penetrating action of electromagnetic waves inside solid bodies, the capability to observe inside distributed targets is precluded. Under these conditions, imaging action is provided only on the surface of distributed targets. The present work describes an imaging method based on the analysis of micro-movements on the Khnum-Khufu Pyramid, which are usually generated by background seismic waves. The results obtained prove to be very promising, as high-resolution full 3D tomographic imaging of the pyramid's interior and subsurface was achieved. Khnum-Khufu becomes transparent like a crystal when observed in the micro-movement domain. Based on this novelty, we have completely reconstructed internal objects, observing and measuring structures that have never been discovered before. The experimental results are estimated by processing series of SAR images from the second-generation Italian COSMO-SkyMed satellite system, demonstrating the effectiveness of the proposed method.
\end{abstract}


\section{Introduction}
The Pyramid of Khnum-Khufu, also known as the Great Pyramid of Giza or Cheops, is the oldest and largest of the three main pyramids that are part of the necropolis of Giza (Egypt). The infrastructure is built with blocks of granite, weighing approximately 2.5 tons each. To complete the work is estimated to have taken at least two and a half million blocks, put in place with millimeter precision in a short period of time, estimated at around 15 or 30 years \cite{lehner1997complete}. 
The pyramids of Egypt, despite being one of the oldest and largest monuments on Earth, to date there is still no common and scientifically established idea on how they were built \cite{hawass1998pyramid,smith2018great}.
The Red Sea was the most important Harbor Facilities at the time of King Khufu \cite{tallet2016harbor}, where an exceptionally well-preserved harbor complex from the Early Old Kingdom at Wadi al-Jarf along the Egyptian coast of the Red Sea has been excavated. 

In studying the origin of the pyramids, we believe we should not overlook the existence of ancient mythological writings.
A study concerning the myths and folklore of the ancient peoples of the world, highlighting all the similarities between them, was made in \cite{bhatt2019subcontinent}. The argument that myths are insignificant, often considered mere stories passed on through generations has been challenged. The authors are open to the possibility that a technologically more advanced civilization existed before a known timeline, where the existence of various glacial ages \cite{kump2005foreshadowing} prevented the passing down of history. Focusing on the mythical cities mentioned in ancient Indian texts, describing how that subcontinent was an integral part of this \cite{bergendorff2019social,noc2015analyse}.
However, how the Egyptian Pyramids were built has remained an enduring mystery\cite{houdin2007construction,baud2015djeser,dodson2000layer}. A theory that the pyramids were cast of cement-like conglomerate made directly in-situ using granular limestone aggregates and an alkali-silicate binder, is proposed in \cite{barsoum2006microstructural}, and evidences are also discussed in \cite{harrell1993great, smyth1877our}.
In order to obtain an accurate perception of how the pyramids were constructed, various engineering hypotheses were evaluated in \cite{tasellari2013great}, noting that in their current form they lay the foundations for new theories. 
At present the general academic consensus is that pyramids served as funerary monuments and burial sites for the pharaohs. However, it is also widely
theorized that such infrastructures may have been built for another purpose. On an aseptic panoramic view, many connections can be found between the pyramids, vibrations and many mechanical devices reminiscent of hydraulic systems, resonance chambers and acoustic filters \cite{malanga1997cheope,hancock2011fingerprints}. The energy to
make the pyramids vibrate can be provided by natural environment and the Earth's atmosphere infrasound vibrations may provide the source of such energy \cite{bedard2000atmospheric}. This provides a basis for the discussion of special classes of waves, including mountain Lee-waves, infrasound, progressive waves in the lower atmosphere, and waves in the upper atmosphere and ionosphere \cite{gossard1975waves}. Atmospheric sound models and extended to the combination effects of both finite depth ocean and source directivity in both elevation and azimuth angles is studied in \cite{de2020atmospheric}.

Acoustic waves are widely used in the field of archaeology. The acoustics of three important World Heritage sites such as the five caves in Spain, the Stonehenge stone circle in England and the Paphos Theatre in Cyprus are studied in \cite{till2019sound}. Groundwater can influence the geomagnetic field measured in the subsurface. The level of water in the rock determines its electrical conductivity, and thus changes the magnitude of the telluric currents induced in the rock by the change in magnetic fields generated in the ionosphere. This can be studied by using several magnetometers at different points in the subsurface. Geomagnetic signals using two magnetometers were successfully monitored in \cite{henry2014monitoring} by setting an optimal electrokinetic magnitude signal upper-bound.
A methodology used for self-potential and seismic-electromagnetic measurements, both for on-site and laboratory experiments as well as for modeling is extensively described in \cite{jouniaux2012electrokinetics}. The research also provides the bibliography on studies carried out in hydrology to remotely detect water flows, to deduce their thickness, and to predict their hydraulic conductivity. The observation method discussed also proposes the detection of fractures in wells, which is also useful in trying to study earthquakes.
Recent theoretical and experimental studies have produced several unusual and interesting results on the cold fusion of matter experienced on dense lithium \cite{guillaume2011cold}. The existence of this exciting propriety of matter relates to zero-point energy estimates that suggest quantum effects play a significant role in shaping the phase diagram of lithium.
The vibration-induced property change in the melting and solidifying process of silver nanoparticles with the use of molecular dynamics simulation was found in \cite{zheng2017vibration}. 

The general problem of acoustic wave propagation through parallel paths is addressed by the information theory of two-port telecommunication networks. This allows any mechanical system to be considered as a single element with two gates. Such a circuit is schematically represented with concentrated elements by admittance that have a value compared to the sum of the corresponding instantaneous admittance existing along the parallel paths. In order to calculate all parameters including those representing transmission losses (e.g. the standing wave ratio), in the case of a non-adaptive paths, the theory of transmission lines is well applicable. A Quincke-tube acoustic filter therefore uses two parallel paths. In order to optimize the maximum wave propagation through the system, the Quincke-tube must be adapted and, through the choice of length and thickness of the ducts, can produce a selective transmission loss, so that it can operate as an acoustic filter. In \cite{hixson1963quincke}, the transmission loss characteristics of several other variations in duct sizes and lengths are presented, together with some very limited experimental data. The phenomenon of pressure pulsations in pipeline systems caused by centrifugal pumps or reciprocating compressors are known to have detrimental effects on industrial applications. An experimental investigation of the attenuation mechanism of a Herschel-Quincke device and its effectiveness in damping pressure pulsations when applied to a resonant piping system had been presented in \cite{lato2019passive}.

The development of mechanical engineering in ancient Egypt through the stone industry was
described in \cite{hassaan2016mechanical}, covering the period from the Predynastic to the Old Kingdom. The characteristics and innovations of stone vessels available in these periods were analyzed.
Conventional sound absorbers can hardly possess the good performance of low-frequency and broadband absorption simultaneously. In order to combine these two functions into one kind of absorbers, the gradually perforated porous materials backed with Helmholtz resonant cavity are proposed in \cite{liu2021gradually}.
A strategy to design three-dimensional elastic periodic structures endowed with complete band-gaps, the first of which is ultra-wide, where the top limits of the first two band-gaps are overstepped in terms of wave transmission in the finite structure is proposed in \cite{d2017mechanical}. Thus, subsequent band-gaps are merged, approaching the behavior of a three-dimensional low-pass mechanical filter.

The debate on how the granite blocks could have been transported up the full height of the pyramids is still an open one. To this end, the theory of in-situ formation of the blocks by means of a cement mixture has also been formulated. Most synthetic stones can be made from re-agglomerated materials. Starting with a mineral substance such as granite rock or naturally eroded, disintegrated or not-aggregated limestone, it is given a compact structure using a binder, such as a geological glue that agglomerates to bond the mineral particles to each other. The result is a new rock with the same mechanical characteristics as a natural equivalent. Such a technique is supposed to have been used to build the pyramid of Khnum-Khufu \cite{davidovits2009pharaohs, barsoum2006microstructural}.

It has been also studied that the generation of waves of any nature can be also correlated with the brain waves of the living creatures. Previous correlations between geomagnetic activity and quantitative changes in electroencephalographic power revealed interesting associations with the right parietal lobe, compatible with theta activity, and the right frontal region for activity in gamma frequencies. During the experiment \cite{mulligan2012experimental}, subjects were exposed to no magnetic field first, then they were exposed to a purely magnetic field of 20 nT or 70 nT, at a frequency of 7 Hz, with amplitude-modulated signals for 30 minutes. Quantitative electroencephalographic (QEEG) measurements were taken before, during and after exposure and results support the thesis that magnetic field fluctuations are primarily responsible for the significant geomagnetic-QEEG correlations reported in several studies. Studying the effects of weak complex magnetic fields on the neuroplasticity of rats, following the induction of early epilepsy, an unprecedented increase in post-crisis mortality (76\%) was observed in young rats that had been exposed perinatally to magnetic fields of 7 Hz with a maximum intensity of about 5 nT.  Rats exposed to fields less intense or more intense than this frequency did not show this magnitude of significant mortality \cite{st2007enhanced,whissell2007developmental}.

Back to the pyramid of Khnum-Khufu, for over a century it has been known that the beams forming the ceiling of the King's Chamber and those of the first and second Relieving Chambers in the Great Pyramid are cracked. However, the temporal origin of these cracks is still unknown. The results of a 3D virtual reality computer simulation designed to determine precisely when the beams cracked are reported in \cite{10.2307/24555438}.
Several 3D imaging techniques applied on the Khnum-Khufu pyramid are developed in \cite{bui2011imaging}. Among all the theories formulated to try to explain how the pyramid of Khnum-Khufu was built, there is also the hypothesis of the existence of an internal ramp that goes around the pyramid several times.  This theory could prove the fact that the pyramid could have been built in twenty years \cite{lheureux2010analyse}.
Microgravity surveys of the Khnum-Khufu pyramid have shown, considering also the general structure of the pyramid. A new interpretation technique for endoscopy of large finite bodies has been developed in \cite{bui1988application}.
In order to carry out non-invasive internal scans of the pyramid \cite{ivashov2021proposed}, using electromagnetic waves, it is necessary to use special georadar \cite{yoshimura1987non}, which however have the limitation of having little penetration inside the granite.

The study of ancient Egyptian monuments attracted the attention of experts all over the world. A recent event that confirms this is the discovery, using muon sensors  \cite{bross2022tomographic,aly2022simulation}, of the presence of a previously unknown cavity located inside the pyramid of Khnum-Khufu. Since this discovery cannot be directly confirmed by drilling, another independent non-destructive method is needed to confirm this discovery and provide an accurate determination of the location and shape of the cavity. A possible holographic radar simulation framework for the detection of openings or other unknown structures of interest is analyzed in \cite{ivashov2021proposed}.

Research \cite{alvarez1970search} for the first time investigated the possibility to use cosmic-ray detectors involved their ability to measure the angle of arrival of penetrating cosmic rays muons with great precision over a large sensitive area. In \cite{morishima2017discovery} authors reported the discovery of a large void (with a cross-section similar to that of the Grand Gallery and a minimum length of 30 metres)
situated above the Grand Gallery.

The investigation of the microgravimetric measurements on the side of a pyramid could also map the recently discovered “muon chamber” in the Great Pyramid of Khnum-Khufu in Egypt. In\cite{pavsteka2022discovery} the exploitation of technical capabilities of modern gravimeters, is  used to perform three-dimensional model calculations with realistic model parameters. A gradiometer survey has been carried out in  \cite{odah2005gradiometer} over a surface area of 100m $\times$ 100m to achieve the purpose and the magnetic data were processed using Geoplot software in order to obtain high quality images of hidden structures inside the Khnum-Khufu pyramid. The results obtained show the presence of interconnected large tomb structures composed of mud-bricks; some other ancient rooms and walls are also present.

A climbing robot called “Djedi” has been designed, constructed, and deployed in \cite{richardson2013djedi} to explore shafts of the Queen's chamber within the Great Pyramid. The Djedi robot is based on the concept of inchworm motion and is capable of carrying a long reach drill or snake camera. The robot successfully climbed the southern shaft of the Great Pyramid, deployed its snake camera, and revealed writing not seen for thousands of years. Robot design, including climbing steps in the shaft and lessons learned from experimental deployment, has been designed in \cite{richardson2013djedi}.

Satellite remote sensing is widely used in the field of archaeology \cite{evans2006use,chen2017overview}.  Data from the use of SAR to survey the southern Maya plains suggest that large areas were continuously drained by ancient canals that may have been used for intensive cultivation. In agreement with the authors of \cite{adams1981radar}, SAR remote sensing confirmed the existence of the canals. Through excavations and in-situ ground surveys, they provided sufficient comparative information. Correlating all the data, it was concluded that the Maya civilization, of the Late Classic period, was firmly based on intensive and large-scale cultivation of marshy areas.

Research \cite{brichieri2020spurred} found an ideal model configuration, associated  with  spiral  ramps, demonstrating how Egyptians could have built the pyramids. 
In the past, Synthetic Aperture Radar (SAR) vibrations have been very useful in estimating key vessel characteristics. Research \cite{filippo2019cosmo,biondi2019micro} propose a novel strategy to estimate the micro-motion (mm) of ships from SAR images. The proposed approach is for mm estimation of ships, occupying thousands of pixels, processes the information generated during the coregistration of several re-synthesized time-domain and not overlapped Doppler sub-apertures of COSMO-SkyMed satellite single-look complex (SLC) data.
Authors of \cite{Pia_1} propose a new procedure to monitor critical infrastructures like the Mosul dam, processing COSMO-SkyMed data. The proposed procedure is an in-depth modal assessment based on the mm estimation, through a Doppler sub-apertures tracking and a multi-chromatic analysis.   
The procedure described above was made available to perform a comprehensive survey of large road bridges, according to \cite{chen2018structural}. The authors of \cite{biondi2020perspectives,9224142,8714049} successfully formulated a comprehensive procedure to perform structural health monitoring using SAR. The technique allows to successfully estimate the position and shape of cracks on bridges in order to prevent their collapse.

In this paper we use a new method based on the tomographic reconstruction of mm, with the aim to perform imaging of the principal targets that make up visible the main internal structure of the pyramid. We use the similar methods already experimented in \cite{Pia_1} to search for cracks in large infrastructures, but not for tomography. The physical principle we use is that of estimating the vibrations captured by the Khnum-Khufu pyramid during the SAR observation time interval. The vibration estimation is done by evaluating the Doppler centroid anomalies, an indispensable parameter that is used during the SAR azimuth focusing process. We use Doppler sub-apertures to estimate the vibrations present on the pyramid. The vibration energy is generated from many sources such as wind. Great contribution in terms of vibration energy, is also generated by the city of Cairo, which is located closely to the pyramid of Khnum-Khufu and by the presence of Nile river.

We processed several SAR images observed in the Vertical-Vertical (VV) polarization, and the estimated mm allows us to visualize the principal internal components present in the pyramid.
We can state that the experimental results we propose definitively solve one of the oldest mysteries of human existence, the complete solution of the internal structure of Khnum-Khufu. 
To this end, in order to provide a more complete contribution to our work, we have firstly investigated the details of the external structure of all the pyramids belonging to the Giza Plateau (Khufu, Kefren and Menkaure), then we concentrated on studying the internal structure of the pyramid of Khnum-Khufu alone, providing a complete and detailed 3D reconstruction of all the known and unknown chambers, based on tomographic SAR measurements. 
In the paper, we provide a complete list of the internal structures measured by tomography, each of them marked with a unique sequential number.

\begin{figure} 
	\centering
	\includegraphics[width=12.0cm,height=7.5cm]{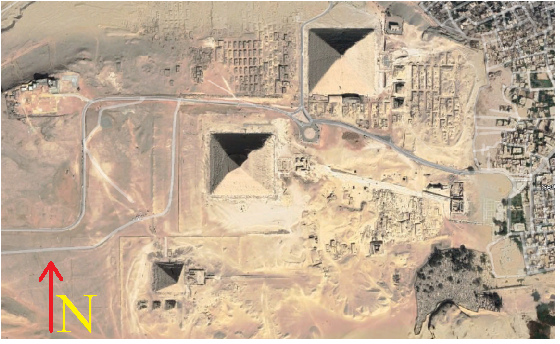}
	\caption{Electro-optical satellite representation of the Giza plateau. The pyramidal infrastructure is visible, oriented to the North.}
	\label{GIZA-Plateau_3}
\end{figure}

\begin{figure} 
	\centering
	\centering
	\includegraphics[width=15.0cm,height=6.0cm]{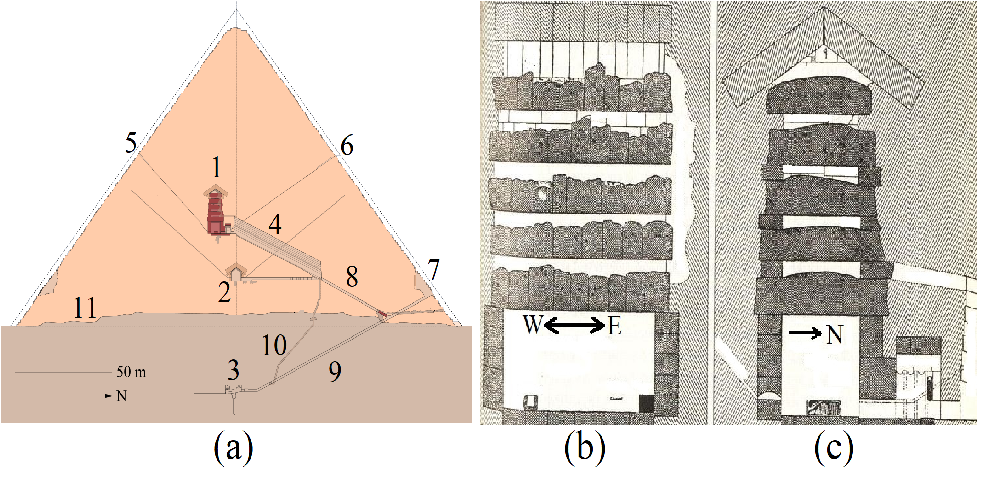}
	\caption{Khnum-Khufu scheme of the known internal structures.}\label{Knum_Khufu_1}
\end{figure}

\begin{table}[tb!]
	\caption{Characteristics of the SAR acquisitions.}
	\begin{center}
		\begin{tabular}{ p{3cm} p{3cm}}
			\hline 
			SAR parametrer & Value\\
			\hline
			Chirp bandwidth& 250 MHz\\
			PRF& 2 kHz\\
			PRT& 0.23 ms\\
			Antenna length & 6 m\\
			Type of acquisition & Stripmap\\
			Polarization & HH\\
			Acquisition duration & 5 s\\
			Platform velocity & 7 km/s\\
			Observation height & 650000 m\\
			\hline
		\end{tabular}
		\label{Table_1}
	\end{center}
\end{table}
\section{Giza Plateau Presentation and Description}
The pyramid of Khnum-Khufu is a monumental structure built mainly of granite blocks, its orientation is almost perfectly aligned to the North. The monumental complex of the Giza plateau is represented in Figure \ref{GIZA-Plateau_3}. The three pyramids, Khnum-Khufu (top right), Kefren (located in the center) and Menkaure (the last on the bottom left) can be observed. In this context our work focuses on visualizing the vibrational tomographic profile of the pyramid of Khnum-Khufu. The Figure \ref{Knum_Khufu_1} (a) is the schematic representation of the North-south central section of the infrastructure. The figure represents the schematic of what is known and the main parts of the infrastructure are numbered sequentially from 1 to 11. The object consists of the Zed and the King's chamber, with its sarcophagus inside. The Zed, the details of which can be seen in Figure \ref{Knum_Khufu_1} (b) and (c), is a large monument made entirely of granite, consisting of an upper roof made of two oblique granite slabs, and five parallel stone slabs, spaced at varying distances from each other. Each stone has its upper face not smooth, so each surface has a pronounced roughness. On the contrary, each of its lower faces is extremely smooth. Below this monument is the King's room. Boh the Zed and the king's room are off-axis with respect to the apex of the pyramid and are located toward the south on the north-south symmetry plane. Object 2 of Figure \ref{Knum_Khufu_1} (a), is the Queen's room, a smaller volume object located on the axis of the pyramid and below the King's room. As can be seen from the figure \ref{Knum_Khufu_1} (a), Object 2, unlike the King's room, is located exactly under the apex of the pyramid. The last room is Object 3 which is also off-axis of the pyramid, it is also shifted to the south, but in this case, it is located underground. It is usually called the unfinished room. Object 4 represents a large corridor that connects the King's room with the Queen's room, it is called the Grand Gallery. Objects 5, and 6 are air ducts, while the remaining ducts 8, 9, and 10 connect the Grand Gallery with the Queen's room and the unfinished room (the one located below ground). Object 7 is the entrance to the pyramid, and finally, line 11 indicates the surface on which the pyramid sits.

\begin{figure} 
	\centering
		\centering
		\includegraphics[width=8.0cm,height=6.0cm]{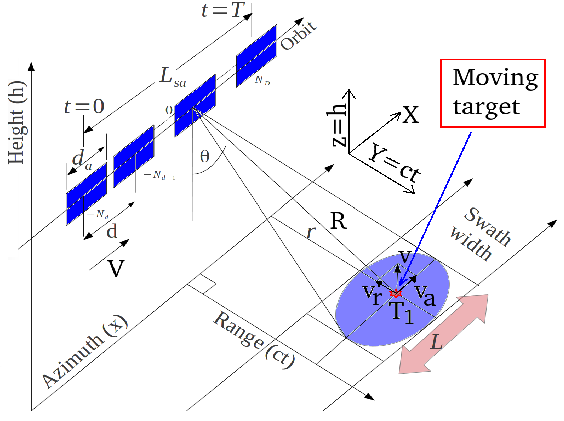}
	\caption{SAR acquisition geometry.}\label{Geometry_1}
\end{figure}

\section{Methodology}\label{Methodology}
In this work, the m-m technique is used to perform sonic imaging by processing a single synthetic aperture radar (SAR) image in the single-look-complex (SLC) configuration. The technique involves the m-m estimation belonging to the Khnum-Khufu pyramid and is generated by the background ripple underground seismic activity that reflects superficial vibrations. The m-m estimation is done through MCA, performed in the Doppler direction. Multiple Doppler sub-apertures, SAR images with lower azimuth resolution, are generated to estimate the vibrational trend of some pixels of interest. The infra-chromatic displacement is calculated through the pixel tracking technique \cite{filippo2019cosmo,biondi2019micro}, using high-performance sub-pixel coregistration \cite{biondi2020monitoring,biondi2020perspectives}. Vibrations observed along the tomographic view-direction, embedded into the multi-chromatic Doppler diversity, are focused along the height (or depth) dimension, and developing high-resolution tomographic underground imaging. 

The SAR synthesizes the electromagnetic image through a ''side looking'' acquisition, according to the observation geometry shown in Figure \ref{Geometry_1}, where:
\begin{itemize}
	\item $r$ is the zero-Doppler distance (constant); 
	\item $R$ is the slant-range;
	\item $R_0$ is the reference range at $t=0$;
	\item $d_a$ is the physical antenna aperture length;
	\item $V$ is the platform velocity;
	\item $d$ is the distance between two range acquisitions;
	\item $G_{sa}$ is the total synthetic aperture length;
	\item $t$ is the acquisition time variable;
	\item $T$ is the observation duration;
	\item $t=0$ and $t=T$ are the start and stop time acquisition respectively;
	\item $L=\frac{\lambda r}{d_a}$ is the azimuth electromagnetic footprint width;
	\item $\theta$ is the incidence angle of the electromagnetic radiation pattern.
\end{itemize}
All the above parameters are related to the staring-spotlight SAR acquisition that is adopted in this work. The SAR data belonging to the electromagnetic image are formed through the focusing process that involves the application of a two-dimensional matched filter acting in the range direction and in the azimuth direction. 
The SLC signal resulting from compression is given by \cite{john1991synthetic}:
\begin {eqnarray}\label{Eq_6}
\begin{split}
	&s_{SLC}(k,x)=2N\tau \exp\left[-\jmath\frac{4\pi}{\lambda}r\right] \textrm{sinc}\left[\pi B_{c_r}\left(k-\frac{2R}{c}\right)\right] \textrm{sinc}\left[\pi B_{c_D}x\right] \\
	&\textrm{for} \ x=kt, \ k=\{0,1,\dots,N-1\} , \ x=\{0,1,\dots,M-1\}, \ \textrm{with} \ N,M \ \in \mathbb{N}.
\end{split}
\end {eqnarray}

Equation \eqref{Eq_6} represents the focused SAR signal generated by the back-scattered electromagnetic energy of a point target supposed to be stationary. The terms $ B_{c_r}$, and $B_{c_D}=\frac{4 Nd}{\lambda r}$ are the total chirp and Doppler bandwidths respectively. The total synthetic aperture is equal to $L_{sa}=2Nd$ and the azimuth resolution $\delta_D\approx \frac{1}{B_{c_D}}=\frac{\lambda R}{2L_{sa}}$.
In \eqref{Eq_6} the $\frac{2\vec{\boldsymbol{R}}}{c}$ parameter identifies the position in range where the maximum of the sinc function is positioned, while in azimuth it is centered around ''zero''.
In the case where the peak of the sinc function has a nonzero coordinate along the azimuth dimension, Equation \eqref{Eq_6} can be recast as:

\begin {eqnarray}\label{Eq_8}
\begin{split}
	&s_{SLC}(k,x)=2N\tau \exp\left[-\jmath\frac{4\pi r}{\lambda}\right] \textrm{sinc}\left[\pi B_{c_r}\left(k-L_{c_g}\right)\right] \textrm{sinc}\left[\pi B_{c_D}\left(x-L_{D_h}\right)\right] \\
	&\textrm{for} \ L_{c_g}, \ L_{D_h} \ \in \mathbb{N},
\end{split}
\end {eqnarray}
where the DFT is equal to:

\begin {eqnarray}\label{Eq_8_DFT}
\begin{split}
	S_{SLC_{F}}(n,q)&=DFT2\left\lbrace2N\tau \exp\left[-\jmath\frac{4\pi r}{\lambda}\right] \textrm{sinc}\left[\pi B_{c_{r}}k\right] \textrm{sinc}\left[\pi B_{c_D}x\right]\right\rbrace \\
	&=2N \tau\exp\left[-\jmath\frac{4\pi r}{\lambda}\right]\sum_{k=0}^{N-1}\sum_{x=0}^{M-1} \textrm{sinc}\left[\pi B_{c_{r}}n\right] \textrm{sinc}\left[\pi B_{c_D}q\right]\\
	&\exp\left(-\jmath\frac{2\pi k n}{N}\right)\exp\left(-\jmath\frac{2\pi x q}{M}\right)\\
	&=2N \tau\exp\left[-\jmath\frac{4\pi r}{\lambda}\right]\frac{1}{\pi B_{c_{r}}}\textrm{rect}\left[\frac{n}{\pi B_{c_{r}}}\right]\frac{1}{\pi B_{c_D}} \textrm{rect}\left[\frac{q}{\pi B_{c_D}}  \right]\\
	&\exp\left(-\jmath 2 \pi n L_{c_g}\right)\exp\left(-\jmath 2 \pi q L_{D_h}\right),
\end{split}
\end {eqnarray}
which has a rectangular shape.
\subsection{Doppler Sub-Apertures Model}\label{Doppler_Sub_Apertures}
\begin{figure} 
	\centering
	\begin{subfigure}
		\centering
		\includegraphics[width=15.0cm,height=4.0cm]{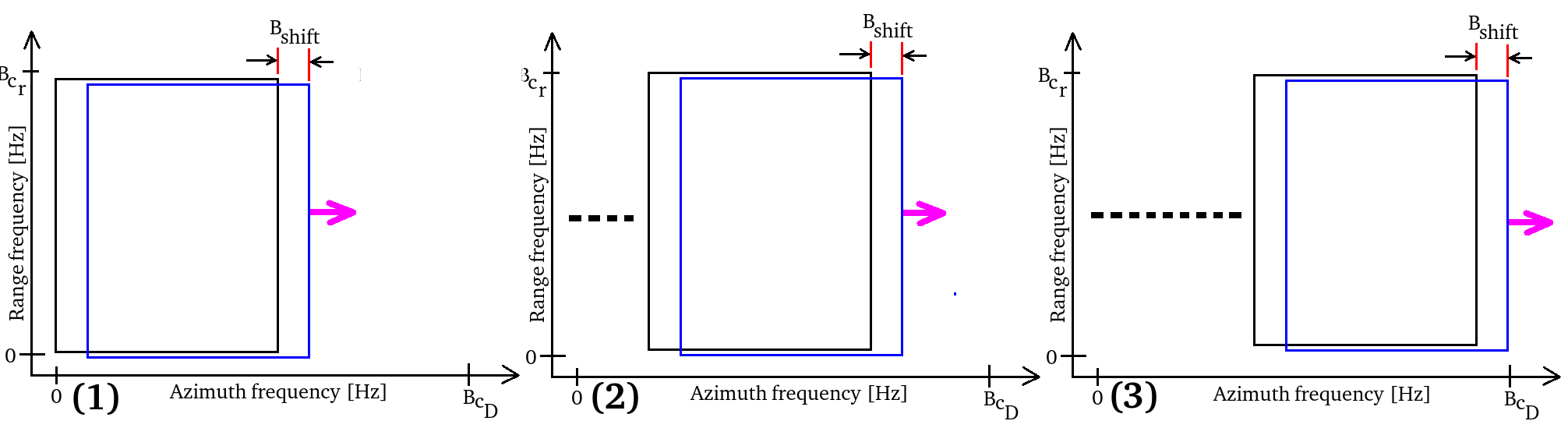}
	\end{subfigure}\quad
	\caption{Doppler sub-aperture strategy.}\label{Bandwidth_Strategy_1}
\end{figure}

In this paper we experiment a strategy that employs Doppler sub-apertures, that are generated to measure target motion. Figure \ref{Bandwidth_Strategy_1} represents the used bandwidth allocation strategy. From the single SAR image we calculate the 2D digital Fourier Transform (DFT) which, according to \eqref{Eq_8_DFT}, has a rectangular shape. As can be seen from Figure \ref{Bandwidth_Strategy_1}, $B_{C_D}$ is the total Doppler band synthesized with the SAR observation, while $B_{D_L}=\frac{B_{C_D}}{2}$ is the bandwidth we left out from the matched-filter boundaries, to obtain a sufficient sensitivity to estimate target motions. In this context formula \ref{Eq_8_DFT} is the focused SAR spectrum, at maximum resolution, thus exploiting the whole band $\{B_{c_r},B_{C_D}\}$, in accordance with the frequency allocation strategy shown in Figure \ref{Bandwidth_Strategy_1}, the following range-Doppler sub-apertures large-matrix is constructed for the Master multi-dimensional information:

\begin {eqnarray}\label{Eq_8_matrix}
\begin{split}
	&S_{SLC}(k,x)_{M}= \begin{bmatrix}
		S_{SLC}(k,x)_{M_{\{1,1\}}} & S_{SLC}(k,x)_{M_{\{1,2\}}} & S_{SLC}(k,x)_{M_{\{1,3\}}} & \dots & S_{SLC}(k,x)_{M_{\{1,N_D\}}}
	\end{bmatrix}\\
	&\textrm{for} \ N_D \ \in \mathbb{N},
\end{split}
\end {eqnarray}
and for the slave, the following large-matrix is presented:
\begin {eqnarray}\label{Eq_9_matrix}
\begin{split}
	&S_{SLC}(k,x)_{S}= \begin{bmatrix}
		S_{SLC}(k,x)_{S_{\{1,1\}}} & S_{SLC}(k,x)_{S_{\{1,2\}}} & S_{SLC}(k,x)_{S_{\{1,3\}}} & \dots & S_{SLC}(k,x)_{S_{\{1,N_D\}}}
	\end{bmatrix}\\
	&\textrm{for} \ N_D \ \in \mathbb{N},
\end{split}
\end {eqnarray}

The explanation of the chirp-Doppler sub-aperture strategy, represented in Figure \ref{Bandwidth_Strategy_1} is the following: We consider Figure \ref{Bandwidth_Strategy_1}, where Master and slave sub-bands are generated by focusing the SAR image, where the matched-filter is set to exploit a range-azimuth bandwidth equal to ${B_{c_r},B_{c_D}-B_{D_L}}$. The not-processed bandwidths $B_{D_L}$ are divided  into $N_D$ equally-distributed bandwidths steps respectively. At this point $N_c$ rigid shifts of the master-slave system are made along the azimuth bandwidth domain, this is made to populate the entire row of \ref{Eq_8_matrix}, and \ref{Eq_9_matrix}. The process is repeated $N_D$ times for each shift in azimuth, in fact, Figures \ref{Bandwidth_Strategy_1} (1), (2) and (3), represent the azimuth frequency variation strategy when the Doppler bandwidth is located at $N_D$. At each Doppler frequency shift $\frac{B_{c_D}-B_{D_L}}{N_D}$ every element of \ref{Eq_8_matrix}, and \ref{Eq_9_matrix} is populated.

\subsection{Doppler Sub-Aperture Strategy}
The decomposition of the SAR data into Doppler sub apertures is formalized in this subsection, which is performed starting from the spectral representation of the focused SAR data.
To this end, notice that the generic $i-$th chirp sub-aperture 2-dimensional DFT of \eqref{Eq_8} is given by:
\begin {eqnarray}\label{Eq_9}
\begin{split}
	&S_{SLC_{F_i}}(n,q)=DFT2\left\lbrace2N\tau \exp\left[-\jmath\frac{4\pi r}{\lambda}\right] \textrm{sinc}\left[\pi B_{c_{r_i}}\left(k-L_{c_g}\right)\right] \textrm{sinc}\left[\pi B_{c_D}\left(x-L_{D_h}\right)\right]\right\rbrace \\
	&=2N \tau\exp\left[-\jmath\frac{4\pi r}{\lambda}\right]\sum_{k=0}^{N-1}\sum_{x=0}^{M-1} \textrm{sinc}\left[\pi B_{c_{r_i}}\left(n-L_{c_g}\right)\right] \textrm{sinc}\left[\pi B_{c_D}\left(q-L_{D_h}\right)\right]\\
	&\exp\left(-\jmath\frac{2\pi k n}{N}\right)\exp\left(-\jmath\frac{2\pi x q}{M}\right)\\
	&=2N \tau\exp\left[-\jmath\frac{4\pi r}{\lambda}\right]\frac{1}{\pi B_{c_{r_i}}}\textrm{rect}\left[\frac{n}{\pi B_{c_{r_i}}}  \right]\frac{1}{\pi B_{c_D}} \textrm{rect}\left[\frac{q}{\pi B_{c_D}}  \right] \exp\left(-\jmath 2 \pi n L_{c_g}\right)\exp\left(-\jmath 2 \pi q L_{D_h}\right). 
\end{split}
\end {eqnarray}
From the last equation, it turns out that a single point stationary target has a two-dimensional rectangular nature with total length proportional to the range-azimuth bandwidths respectively. The phase term $\exp\left(-\jmath 2 \pi n L_{c_g}\right)\exp\left(-\jmath 2 \pi q L_{D_h}\right)$ is due to the sinc function dislocation in range and azimuth when the SLC SAR data are considered.
In the SAR, the movement of a point target with velocity in both range and azimuth direction is immediately warned by the focusing process, resulting in the following anomalies:
\begin{itemize}
	\item azimuth displacement in the presence of target constant range velocity;
	\item azimuth smearing in the presence of target azimuth velocity or target range accelerations;
	\item range-walking phenomenon, visible as range defocusing, in the presence of target range speed, backscattered energy can be detected over one or more range resolution cells.  
\end{itemize}
In practical cases, the backscattered energy from moving targets is distributed over several range-azimuth resolution cells.
As a matter of fact, considering the point-like target $T_1$ (of Figure \ref{Geometry_1}) that is moving with velocity $\vec{\boldsymbol{v}}_t$ whose range-azimuth and acceleration components are $\{v_r,v_a\}$, and  $\{a_r,a_a\}$, respectively, then we can write
\begin {eqnarray}\label{Eq_10}
\begin{split}
	&R^2(t)=(V t-S_a)^2+(R_0-S_r)^2 \ \textrm{with} \  S_r=v_r t+\frac{1}{2}a_r t^2 \ \textrm{and} \ S_a=v_a t+\frac{1}{2}a_a t^2 \\
	&\left|R(t)\right|=\left|R_0-S_r\right|\left\lbrace 1+\frac{\left(Vt-S_a\right)^2}{\left(R_0-S_r\right)^2} \right\rbrace^\frac{1}{2}.\\
\end{split}
\end {eqnarray}
Considering the following Taylor expansion:
\begin {eqnarray}\label{Eq_11}
\begin{split}
	&\left(1+x\right)^\beta\approx1+\beta x\\
\end{split}
\end {eqnarray}
and that $R_0-S_r\approx R_0$, and $\left(V_t-S_a\right)^2 \approx V^2 t^2 - 2V t S_a$, \eqref{Eq_10} can be written in the following form:
\begin {eqnarray}\label{Eq_12}
\begin{split}
	&\left| R(t) \right| = \left\lbrace \left| R_0-S_r \right| + \frac{1}{2}\frac{\left(Vt-S_a\right)^2}{\left(R_0-S_r\right)} \right\rbrace = |R_0-S_r|+\frac{V^2t^2}{2R_0} \left( 1-\frac{2S_a}{Vt} \right) 
\end{split}
\end {eqnarray}
\begin {eqnarray}\label{Eq_13}
\begin{split}
	&=R_0-S_r+\frac{V^2t^2}{2R_0}-\frac{V t S_a}{R_0} \\
	&=R_0 - v_r t-\frac{1}{2}a_r t^2+\frac{V^2t^2}{2R_0}-\frac{V t \left(v_a t+\frac{1}{2}a_a t^2\right)}{R_0}\\
	&=R_0 - v_r t-\frac{1}{2}a_r t^2+\frac{V^2t^2}{2R_0}-\frac{V v_a t^2}{R_0}-\frac{V a_a t^3}{2R_0}.
\end{split}
\end {eqnarray}
The term $\frac{V a_a t^3}{2R_0}$ can be neglected and by approximating $\left(V^2-2Vv_a\right)\approx\left(V-v_a\right)^2$ Equation \eqref{Eq_13} can be written like:
\begin {eqnarray}\label{Eq_14}
\begin{split}
	&\left|R(Vt)\right|=R_0-v_r t + \frac{t^2}{2R_0} \left[ \left( V-v_a \right)^2 -R_0 a_r\right].
\end{split}
\end {eqnarray}
recasting \eqref{Eq_14} in terms of $x=Vt$, we obtain \cite{4103740}:
\begin {eqnarray}\label{Eq_15}
\begin{split}
	&\left|R(x)\right|=R_0-\epsilon_{r_1}x+ \left[ \left( 1-\epsilon_{c_1} \right)^2 - \epsilon_{r_2} \right] \frac{x^2}{2R_0}, \ x=Vt.
\end{split}
\end {eqnarray}
where:
\begin{itemize}
	\item $\epsilon_{r_1}=\frac{v_r}{V}$ (due to range velocity);
	\item $\epsilon_{r_2}=\frac{a_rR_0}{V^2}$ (due to range acceleration);
	\item $\epsilon_{c_1}=\frac{v_c}{V}$ (due to azimuth velocity).  
\end{itemize}
Thus, the above terms modify the received signal, as shown in \cite{4103740}, and should be taken into account in Equation \eqref{Eq_9} .

\section{Tomographic Model}\label{Methods}
\begin{figure} 
	\centering
	\begin{subfigure}
		\centering
		\includegraphics[width=15.0cm,height=4.5cm]{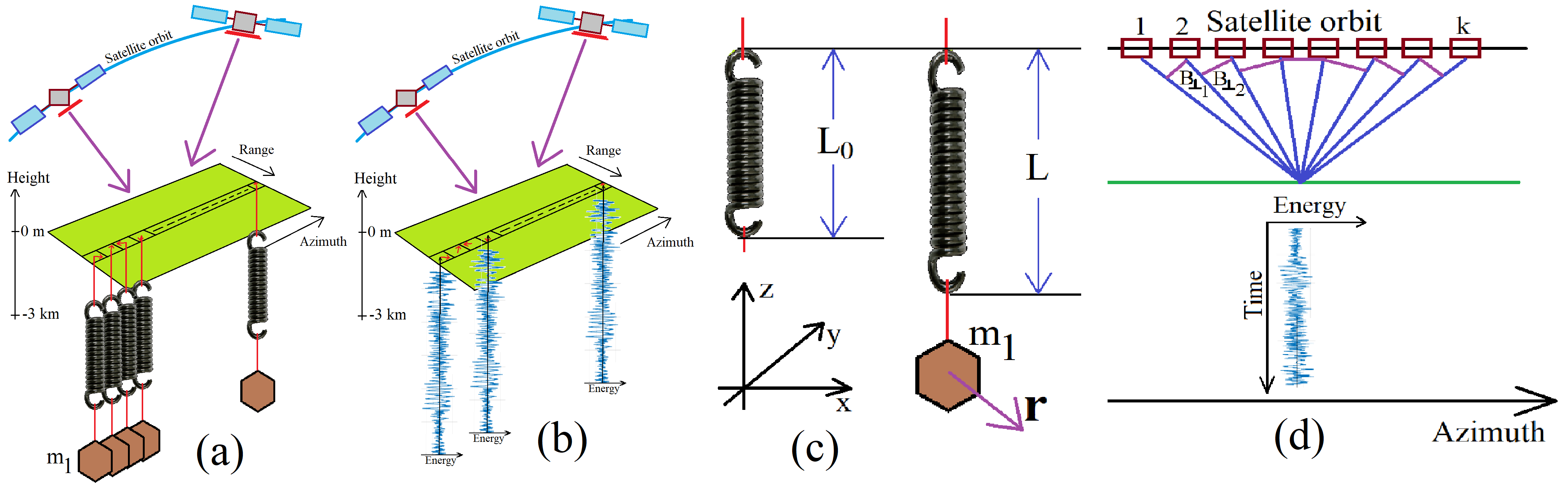}
	\end{subfigure}\quad
	\caption{Tomographic acquisition geometry.}\label{Tomo_1}
\end{figure}

Considering a single SLC image from which we applied the MCA according to the frequency allocation strategy depicted in Figure \ref{Bandwidth_Strategy_1}, the tomogram represented by the line of contiguous pixels shown in Figure \ref{Tomo_1} is calculated. The vibrations present on the tomographic plane extending from the Earth's surface to a depth of a few kilometres is assessed. The figure represents a series of harmonic oscillators anchored on each pixel of the tomographic line, symbolically represented as a spring linked to a mass and oscillating due to the application of harmonic vibrations. Each wave generated by each harmonic oscillator bounces off the surface of the Earth as there is an abrupt variation in the density of the medium (the ground-air boundary). On each pixel a vibrational phasor is observed in time applying Doppler MCA \cite{biondi2020monitoring,biondi2020perspectives}. 
Through the orbital change of view (which is performed in azimuth), an effective subsurface in-depth vibrational scan of the Earth is achieved.

\subsection{Vibrational Model of the Earth}\label{Vibrational_Model_of_cables}
The proposed vibrational model of the Earth surface is schematically shown in Figure \ref{Tomo_1} (a), and (b). The geometrical reference system for both sub-pictures is the range, azimuth and altitude three-dimensional space. For the present case, the vertical dimension represents the depth below the topographic level, (for this specific case the medium boundary is represented by the green plane). The tomographic line of interest is constituted of the series of contiguous pixels laying on the green plane. As can be seen from Figure \ref{Tomo_1} (a) on each pixel belonging to the tomographic line, a mass is hanging using a spring. This system is now induced to oscillate harmonically, helped by the Earth magma instability. These oscillations are schematized as the vibration energy function visible in Figure \ref{Tomo_1} (b). In this context the radar instantaneously perceives this coherent harmonic oscillation. In a mathematical point of view, the Earth displacement is perceived as a complex shift belonging on each pixel of interest. Each instantaneously displacement is estimated between the master image with respect to the slave, where oversimplification shifts are estimated through the pixel tracking technique \cite{biondi2020monitoring,biondi2020perspectives}. The number of tomographic independent looks (depending on the total number of Doppler sub-apertures) are defined by the parameter $k$.

We suppose now the spring being perturbed by an impulsed force. According to this perturbation the rope begins to vibrate describing an harmonic motion (in this context we are not considering any form of friction). Resulting perturbation moves the rope through the space-time in the form of a sinusoidal function. The seismic wave will then reach a constraint end that will cause it to reflect in the opposite direction. The reflected wave will then reach the opposite constraint that will make it reflect in the original direction and returning in the initial location, maintaining the same frequency and amplitude. According to Classical Physics principles, the rebounding wave is superimposed on the arriving wave, and the interference of two sine waves with the same amplitude and frequency propagating in opposite directions leads to the generation of an ideal and perpetual standing wave on the spring. Each vibrational channel is now considered when the spring is able to oscillate into the three-dimensional space, according to specific perturbation nature. When the Earth vibrates, it happens that the length of the spring must also fluctuate. This phenomenon causes oscillations in the tension domain of the spring. It is clear that these oscillations (i.e. the longitudinal ones) propagate through a frequency approximately twice as high as the frequency value of the transverse vibrations. The coupling between the transverse and longitudinal oscillations of the spring can essentially be modeled through non-linear phenomena.
 \begin{figure} 
	\centering
	\includegraphics[width=16.0cm,height=6.5cm]{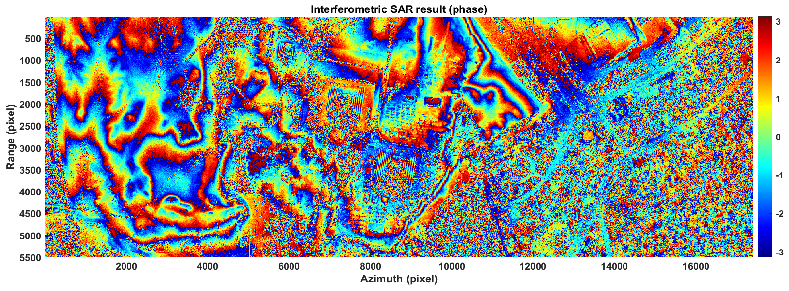}
	\caption{SAR image representing the interferometric repeat-pass phase of the Giza plateau.}
	\label{Interferometric_Phase_1}
\end{figure}
\begin{figure} 
	\centering
	\includegraphics[width=16.0cm,height=6.0cm]{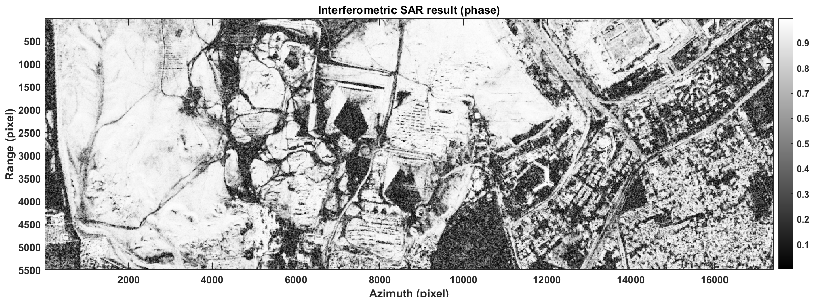}
	\caption{SAR image representing the interferometric repeat-pass coherence of the Giza plateau.}
	\label{Interferometric_Coherence_1}
\end{figure}
{\color{black}Figure \ref{Tomo_1} (c, d) illustrates the oscillating model in the Euclidean space-time coordinates (x,y,z,t), where the satellite motion has been purified from any orbital distortions, so that the geometric parameters used to perform the tomographic focusing can be rigorously understood.} From Figure \ref{Tomo_1} (c), $L$ is the length of the spring when it is at its maximum tension while $L_0$ is its length when no mass is present. Finally the spring has been considered to have an elastic constant equal to $\xi$. The vibrational force applied to the mass $m1$ of Figure \ref{Tomo_1} (c) is equal to \cite{tufillaro1989nonlinear}:
\begin {eqnarray}\label{Eq_16}
\begin{split}
	&F=-4 \xi \mathbf{r} \left(1-\frac{L_0}{\sqrt{L^2+4\mathbf{r}^2}}\right).
\end{split}
\end {eqnarray}
If $\mathbf{r}\ll L$, \eqref{Eq_16} is expanded in the following series:
\begin {eqnarray}\label{Eq_16_Bis}
\begin{split}
	&F=-4 \xi \mathbf{r}(L-L_0)\left(\frac{\mathbf{r}}{L}\right)-8 \xi L_0 \left[\left(\frac{\mathbf{r}}{L}\right)^3-\left(\frac{\mathbf{r}}{L}\right)^5+\dots\right],
\end{split}
\end {eqnarray}
where a precise approximation of \eqref{Eq_16_Bis} is the following cubic restoring force:
\begin {eqnarray}\label{Eq_18}
\begin{split}
	&F=m\ddot{\mathbf{r}}\approx-4 \xi \mathbf{r}(L-L_0)\left(\frac{\mathbf{r}}{L}\right) \left[1+\frac{2L_0}{(L-L_0)}\left(\frac{\mathbf{r}}{L}\right)^2\right].
\end{split}
\end {eqnarray}
Considering \eqref{Eq_18}, the non-linearity dominates when $L \approx L_0$.
If we define:
\begin {eqnarray}\label{Eq_19}
\begin{split}
	&\omega_0=\frac{4 \xi}{m}\left[\frac{\left(L-L_0\right)}{L}\right],
\end{split}
\end {eqnarray}
and
\begin {eqnarray}\label{Eq_20}
\begin{split}
	&\xi=\frac{2L_0}{L^2}\left(L-L_0\right).
\end{split}
\end {eqnarray}
Considering \eqref{Eq_18} we have:
\begin {eqnarray}\label{Eq_21}
\begin{split}
	&\ddot{\mathbf{r}}+\omega^2 \mathbf{r} \left(1+ \xi\mathbf{r}^2\right)=0.
\end{split}
\end {eqnarray}
If we consider damping and forcing \eqref{Eq_21} is modified as:
\begin {eqnarray}\label{Eq_24}
\begin{split}
	&\ddot{\mathbf{r}}+\lambda\dot{\mathbf{r}}+\omega^2\left(1+\xi\mathbf{r}^2\right) \mathbf{r}=\mathbf{f}(\omega t),
\end{split}
\end {eqnarray}
where $\mathbf{f}(\omega t)$ is the forcing term and $\lambda$ is the damping coefficient. 
If non-linearity of \eqref{Eq_24} is sufficiently low, it can be reduced into the following two-degree-of-freedom linear harmonic oscillator:
\begin {eqnarray}\label{Eq_23}
\begin{split}
	&\mathbf{r}(t)=\left(a \cos\omega_0 t, b\sin \omega_0 t\right)\exp\left(\frac{-\lambda t}{2}\right).
\end{split}
\end {eqnarray}
In \ref{Eq_23} $\{a,b\}$ are the instantaneous shifts estimated by the coregistrator. The harmonic oscillator \eqref{Eq_23} is the displacement parameters ${\epsilon_{r_1}, \epsilon_{r_2}, \epsilon_{c_1}}$ estimated by \eqref{Eq_15}.
According to Figure \ref{Tomo_1} (d) the vector representation of $k$ samples of the time-domain function \eqref{Eq_23} consisting in the following multi-frequency data input is considered:
\begin {eqnarray}\label{eq_2}
\mathbf{Y} =\left[\mathbf{y}(1),\dots ,\mathbf{y}(k)\right],\in \mathbf{C}^{k\times 1}.
\end {eqnarray}
The steering matrix $\mathbf{A}(z)=\left[\mathbf{a}(z_{min}),\dots,\mathbf{a}(z_{MAX})\right]$, $\in  	\mathbf{C}^{k\times F}$ contains the phase information of to the Doppler frequency variation of the sub-aperture strategy, associated to a source located at the elevation position $\mathbf{z} \in \{z_{min},z_{MAX}\}$,
\begin{eqnarray}\label{Eq_31}
	\mathbf{A}(\mathbf{K}_z,\mathbf{z})=	\begin{bmatrix}
		1,\exp(\jmath 2\pi k_{z_2} t z_0),\dots,\exp(\jmath 2\pi k_{z_{k-1}} t z_0) \\
		1,\exp(\jmath 2\pi k_{z_2} t z_1),\dots,\exp(\jmath 2\pi k_{z_{k-1}} t z_1) \\
		\dots \\
		1,\exp(\jmath 2\pi k_{z_2} t z_{F-1}),\dots,\exp(\jmath 2\pi k_{z_{k-1}} t z_{F-1})
	\end{bmatrix},
\end{eqnarray}
where $\mathbf{K}_z= \frac{4\pi B_{\perp}}{\lambda \mathbf{r}_i \sin \theta}$, $i=1,\dots,k$, $B_\perp$ is the $i-$th orthogonal baseline which is visible in Figure \ref{Tomo_1} (d), and $\mathbf{r}_i$ is the $i-$th slant-range distance. The standard sonic tomographic model is given by the following relation:
\begin{eqnarray}\label{Eq_35}
	\mathbf{Y}&=\mathbf{A}(\mathbf{K}_z,\mathbf{z}) {\mathbf{h}}(\mathbf{z}).
\end{eqnarray} 
where in \eqref{Eq_35} ${\mathbf{h}}(\mathbf{z}) \in \mathbf{C}^{1 \times F}$, inverting \eqref{Eq_35} I finally find the following tomographic solution:
\begin{eqnarray}\label{Eq_34}
	{\mathbf{h}}(\mathbf{z})&=\mathbf{A}(\mathbf{K}_z,\mathbf{z})^\dagger \mathbf{Y}.
\end{eqnarray} 

In the \eqref{Eq_34} the steering matrix $\mathbf{A}(\mathbf{K}_z,\mathbf{z})$ represents the best approximation of a matrix operator performing the digital Fourier transform (DFT) of $\mathbf{Y}$. The tomographic image $\mathbf{h}(\mathbf{z})$, which represents the spectrum of $\mathbf{A}(\mathbf{K}_z,\mathbf{z})$, is obtained by doing pulse compression.

The tomographic resolution is equal to $\delta_T=\frac{\lambda R}{2 A}$, where $\lambda$ is the sound wavelength over the Earth, $R$ is the slant range, and $A$ is the orbit aperture considered in the tomographic synthesis, in other words, consists to the Doppler bandwidth used to synthesize the sub-apertures. The maximum tomographic resolution obtainable using this SLC data, synthesized at 24 kHz, is as follows. Considering an average speed of propagation of the seismic waves of about $v \approx 6000 \frac{m}{s}$, a frequency of investigation set by us equal to 12500 Hz, the wavelength of these vibrations is equal to about $\lambda=\frac{v}{f}\approx \frac{6000}{12500} \approx 0.48 m$. Considering the above parameters, extending the tomography to the maximum orbital aperture equal to half the total length of the orbit, therefore about $42000$ m, with $R=650000 m$ the tomographic resolution is equal to $\delta_z=\frac{\lambda R}{2 A}= \frac{0.48 \cdot 650000}{2 \cdot 42000} \approx 3.71 m$. This is the tomographic resolution set to calculate all the experimental parts shown in section \ref{Experimental_Rasults_All}.
\begin{figure} 
	\centering
	\includegraphics[width=16.0cm,height=4.5cm]{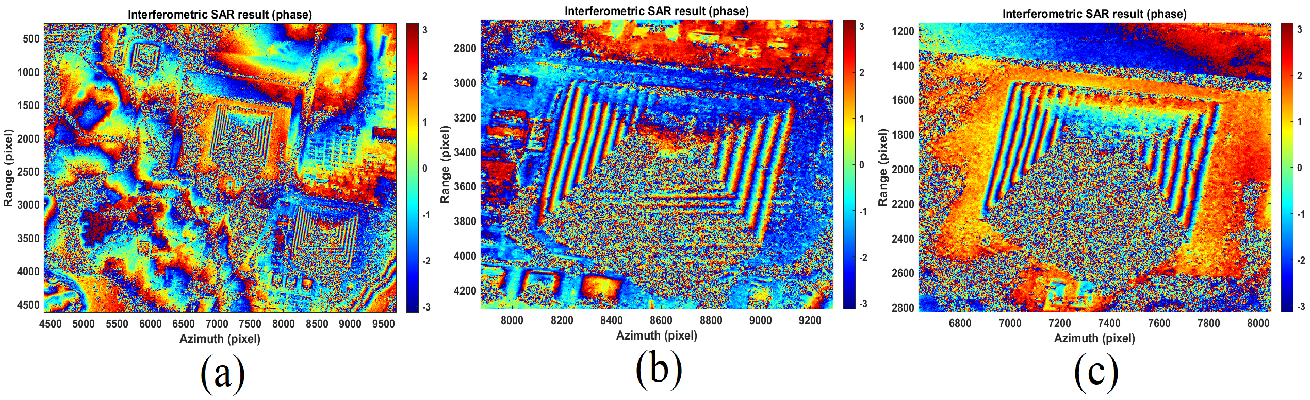}
	\caption{Electro-optical satellite representation of the Giza plateau. The pyramidal infrastructure is visible, oriented to the North.}
	\label{Interferometric_Phase_Cut_1_All}
\end{figure}

\begin{figure} 
	\centering
	\includegraphics[width=16.0cm,height=4.5cm]{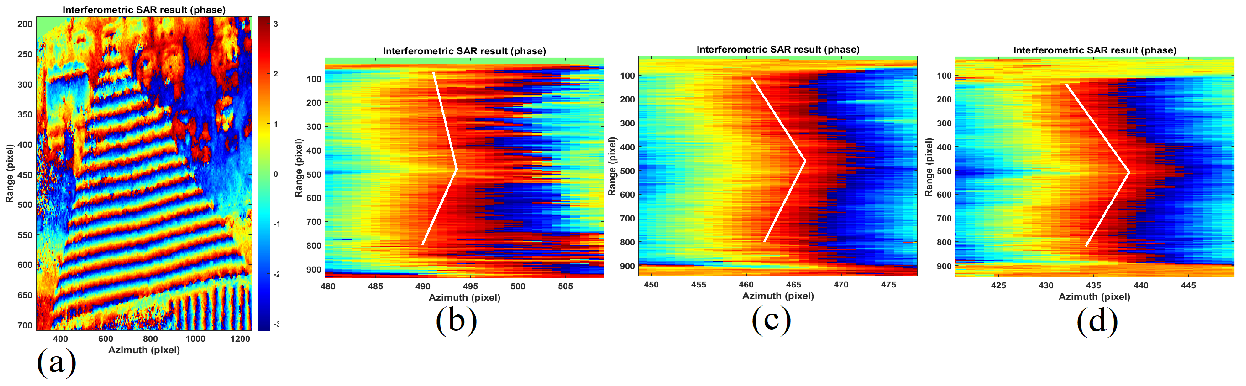}
	\caption{Representation in particular of the pyramid of Khnum-Khufu. Interferometric phases. (a): North side. (b,c,d): magnification of interferometric fringe 1,2, and 3.}
	\label{Intererogramma_3_Khufu_Nord_2}
\end{figure}

\begin{figure} 
	\centering
	\includegraphics[width=16.0cm,height=3.5cm]{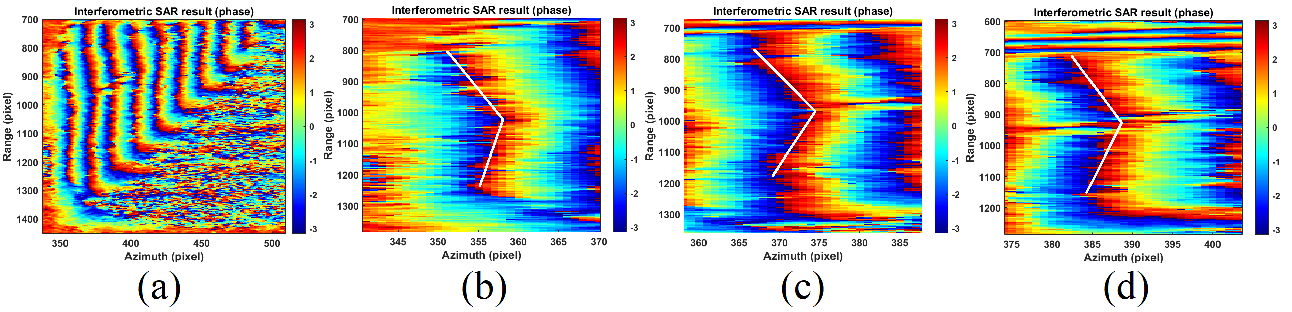}
	\caption{Representation in particular of the pyramid of Khnum-Khufu. Interferometric phases. (a): East side. (b,c,d): magnification of interferometric fringe 1,2, and 3.}
	\label{Intererogramma_3_Khufu_All_Ovest}
\end{figure}

\begin{figure} 
	\centering
	\includegraphics[width=16.0cm,height=4.0cm]{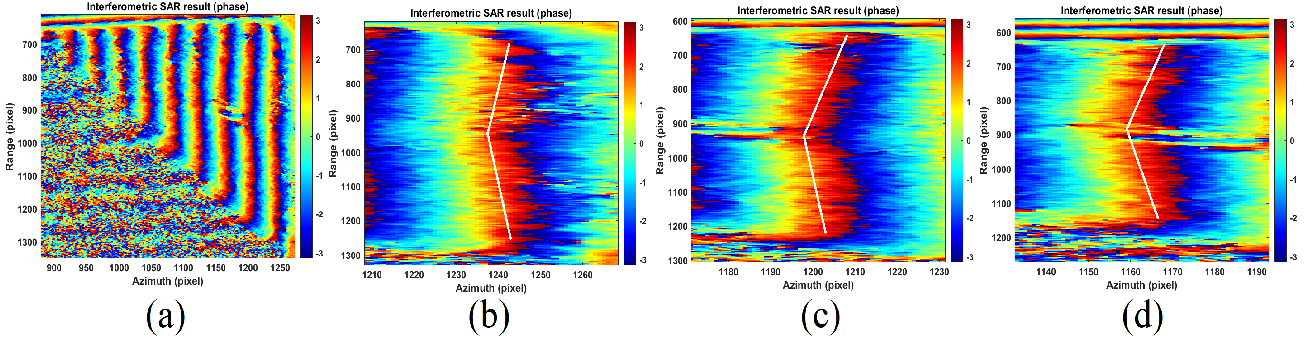}
	\caption{Representation in particular of the pyramid of Khnum-Khufu. Interferometric phases. (a): West side. (b,c,d): magnification of interferometric fringe 1,2, and 3.}
	\label{Intererogramma_3_Khufu_All_Est}
\end{figure}
\begin{figure} 
	\centering
	\includegraphics[width=16.0cm,height=4.0cm]{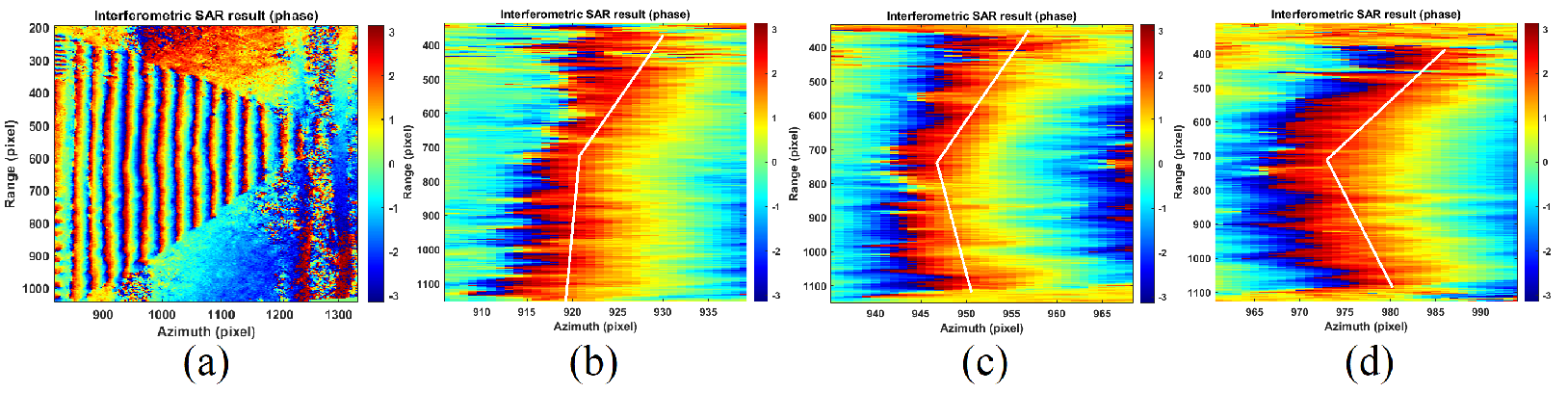}
	\caption{Representation in particular of the pyramid of Kefren. Interferometric phases. (a): North side. (b,c,d): magnification of interferometric fringe 1,2, and 3.}
	\label{Chefren_Nord_All}
\end{figure}

\begin{figure} 
	\centering
	\includegraphics[width=16.0cm,height=4.0cm]{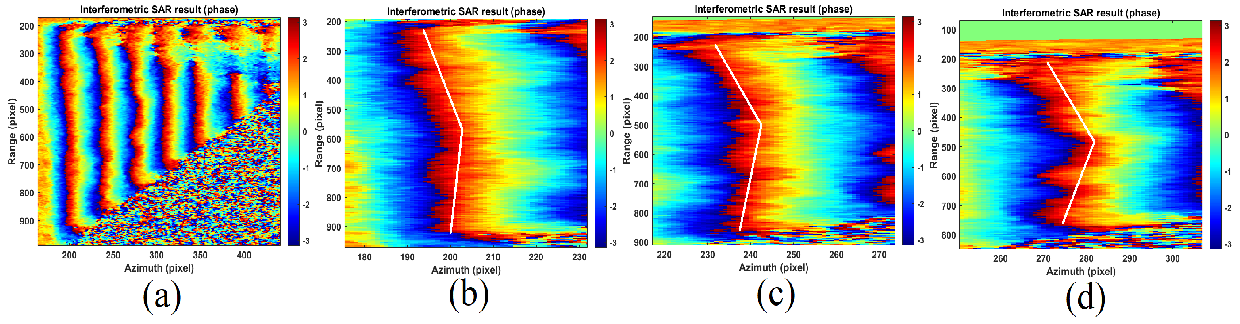}
	\caption{Representation in particular of the pyramid of Kefren. Interferometric phases. (a): Est side. (b,c,d): magnification of interferometric fringe 1,2, and 3.}
	\label{Chefren_Est_All}
\end{figure}
\section{Experimental Results}\label{Experimental_Rasults_All}
In this section we will show all the experimental results that have been made and have been divided into external, and internal, experimental results. In the first case, we show the results provided by SAR interferometry (InSAR). These consist of the evaluation of radar interferometric fringes to demonstrate the actual shapes of the outer facades of all pyramids in the Giza Plateau. The detailed explanation is provided in the subsection \ref{Experimental_Rasults_External}. Similarly the results for the internal vibrational tomography analysis of Khnum-Khufu alone, are discussed in detail in subsection \ref{Experimental_Rasults_Internal}.

\section{External Experimental Results}\label{Experimental_Rasults_External}
This subsection shows and discusses all the results obtained with the aim of revealing new features of the external appearance of all pyramids residing on the Giza Plateau. In order to achieve this goal, we employed the InSAR technique and evaluated the nature of the interferometric fringes and discovered, through the measurement of their inclination, that all pyramids do not each consist of four faces but of eight faces. We found that each face of each pyramid had an inwards bow that became more relevant closer to the ground much like a trough. Figures \ref{Interferometric_Phase_1} show the interferometric fringes generated by two SAR repeat-pass acquisitions, with a suitable spatial baseline in order to generate a series of well-estimated interferometric fringes imprinted on the faces of the pyramids. The interferometric acquisition was performed along a time baseline equal to the complete orbital cycle of the single CSG satellite, which coincides with 16 days. In spite of the substantial number of waiting days, the interferometric acquisition appears not to be very noisy and this quality is confirmed through the evaluation of the coherence parameter whose map is represented in Figure \ref{Interferometric_Coherence_1}. This result appears very good, as a large part of the figure, removing all the areas where the radar shadow is present, maintains coherence levels very close to 1.

The Figure \ref{Interferometric_Phase_Cut_1_All} (a) represent the details of the InSAR fringes measured on the three pyramids (Khufu, Kefren, and Menkaure), while Figure \ref{Interferometric_Phase_Cut_1_All} (b) is the detail of the pyramid of Khnum-Khufu, and finally Figure \ref{Interferometric_Phase_Cut_1_All} (c) is the pyramid of Kefren. Figure \ref{Intererogramma_3_Khufu_Nord_2} (a), (b), (c) and (d), are the particular representations of the SAR interferometric fringes observed on the North face of the Pyramid of Khnum-Khufu (in box (a)). The remaining boxes (b), (c) and (d), represent the first, second and third interferometric fringes observed on the North face of the same pyramid, starting from the bottom. The inclination of the entire face of the pyramid is clearly observed, having symmetry along the height of the geometric figure. The east face of the Khnum-Khufu pyramid is depicted in Figure \ref{Intererogramma_3_Khufu_All_Ovest} (a), while the details of the first, second and third fringe (starting from the ground plane) are shown in Figure \ref{Intererogramma_3_Khufu_All_Ovest} (b), (c) and (d) respectively. Here again, the same effect is observed whereby the single face is divided into two indented half-faces. The West face also presents the same architectural feature, in fact, Figure \ref{Intererogramma_3_Khufu_All_Est} (a), depicts the extension of the interferometric fringes extended over the entire West face, while the details of the first, second and third fringes (always starting from the ground plane), are shown in Figure \ref{Intererogramma_3_Khufu_All_Est} (b), (c) and (d) respectively.

\begin{figure} 
	\centering
	\includegraphics[width=16.0cm,height=4.0cm]{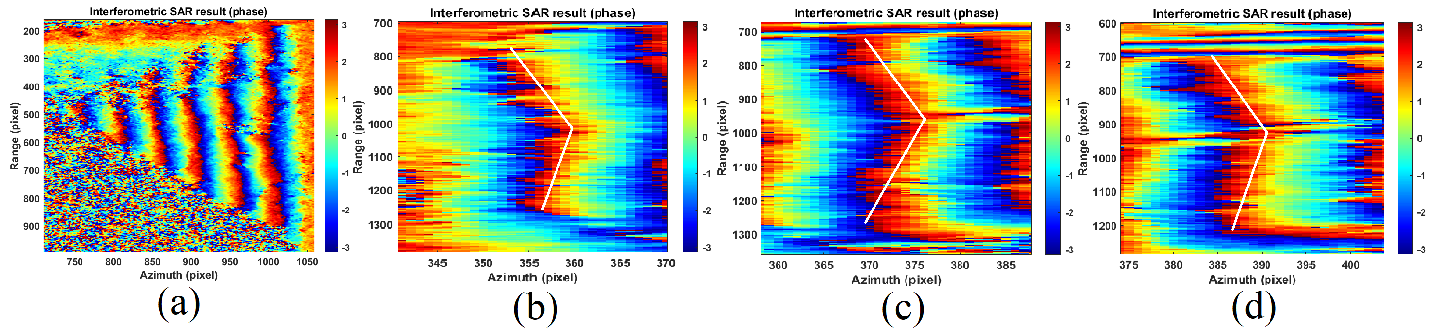}
	\caption{Representation in particular of the pyramid of Kefren. Interferometric phases. (a): West side. (b,c,d): magnification of interferometric fringe 1,2, and 3.}
	\label{Chefren_Ovest_All}
\end{figure}

We now move to the pyramid of Kefren to repeat the same experiments where the same qualitative evaluation of the SAR interferometric fringes present on the three facets of it, except for the South one, will be carried out, reaching the identical conclusions made for the pyramid of Khufu. Figure \ref{Chefren_Nord_All} (a), (b), (c) and (d), are the particular representation of the SAR interferometric fringes observed on the North face of the Pyramid of Kefren (in box (a)). The remaining boxes (b), (c) and (d), represent the first, second and third interferometric fringes observed on the North face of the same pyramid, starting from the bottom. The East Front of the second Kefren pyramid is analyzed in Figure \ref{Chefren_Est_All} (a), depicting the extent of the interferometric fringes as a whole, while the details of the first, second and third fringes (starting from the ground plane) are shown in Figure \ref{Chefren_Est_All} (b), (c) and (d) respectively. The West Façade is also studied in Figure \ref{Chefren_Ovest_All} (a), along with its first, second and third fringe details (starting from the ground plane), which are shown in Figure \ref{Chefren_Ovest_All} (b), (c) and (d) respectively.

\begin{figure} 
	\centering
	\includegraphics[width=16.0cm,height=4.0cm]{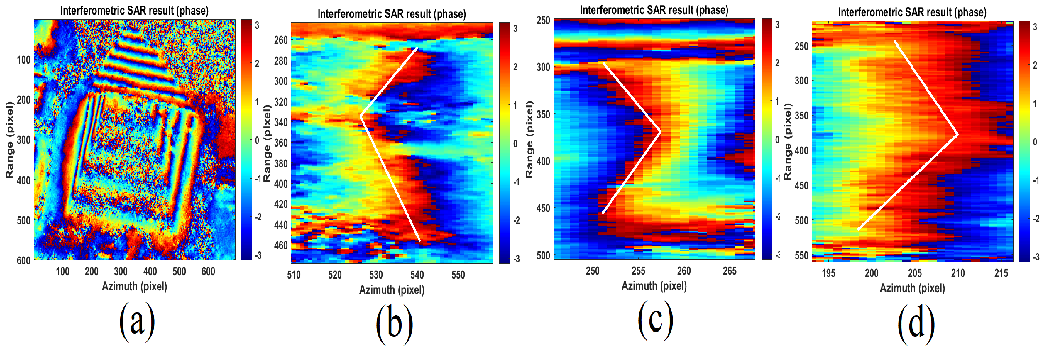}
	\caption{Representation in particular of the pyramid of Menkaure. Interferometric phases. (a): Menkaure pyramid. (b): East-side interferometric fringe particular. (c): West-side interferometric fringe particular. (d): North-side interferometric fringe particular.}
	\label{Micerino_All}
\end{figure}

Although much smaller, the pyramid of Menkaure is also well represented by the interferometric radar. As a matter of fact, Figures (a), (b), (c) and (d) depict the entire pyramid (in Figures (a)), while Figures (b), (c) and (d) depict the first fringe starting from the ground plane of the East, West, and North faces respectively. Surprisingly, the pyramid of Menkaure also consists of eight facets and not four (while maintaining the fact that the south face could not be observed because in radar shadow).
This subsection, which focused on presenting the results of external measurements alone, ends by rigorously demonstrating, through radar measurements, the eight-sided nature of the three pyramids of Khufu, Kefren and Menkaure. In the next subsection, the internal measurements that were made from space will be detailed, in order to carry out for the first time the complete internal mapping of all structures belonging to the pyramid of Khnum-Khufu alone.
\begin{figure} 
	\centering
	\includegraphics[width=10.0cm,height=8.0cm]{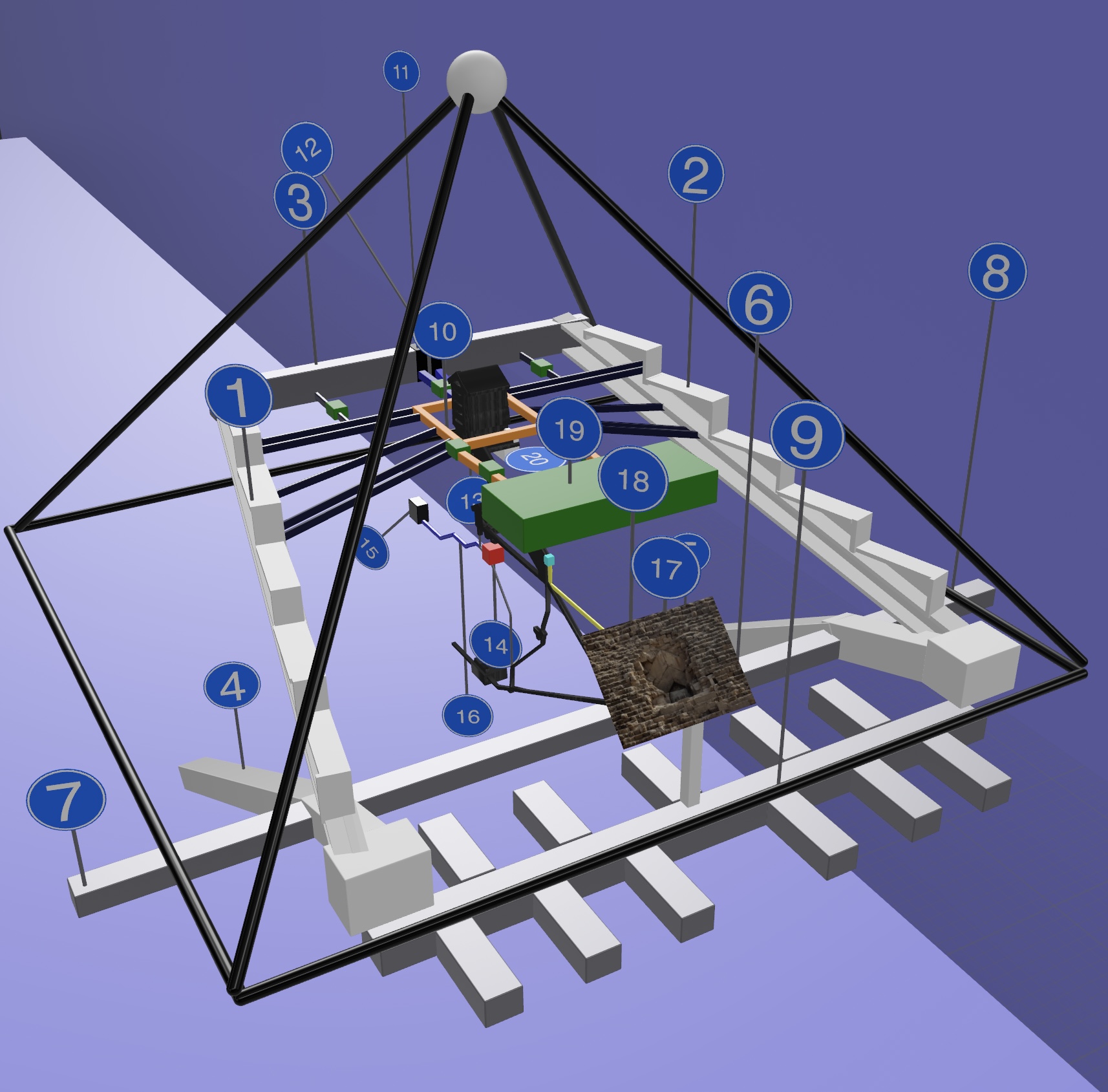}
	\caption{(a): SLC SAR image in magnitude representation of Khnum-Khufu. (b): Schematic representation of Khnum-Khufu.}
	\label{Fully_3D_1}
\end{figure}

\begin{figure} 
	\centering
	\includegraphics[width=15.0cm,height=6.5cm]{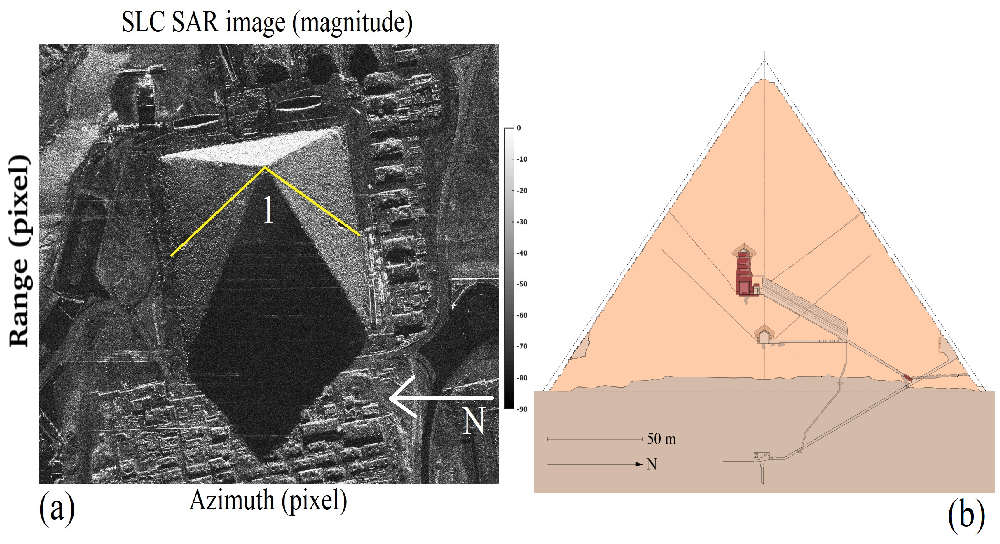}
	\caption{(a): Tomographic map of Khnum-Khufu overlapped to its schematic representation. (b): Tomographic map of Khnum-Khufu.}
	\label{Merged_0}
\end{figure}

\begin{figure} 
	\centering
	\includegraphics[width=15.0cm,height=6.5cm]{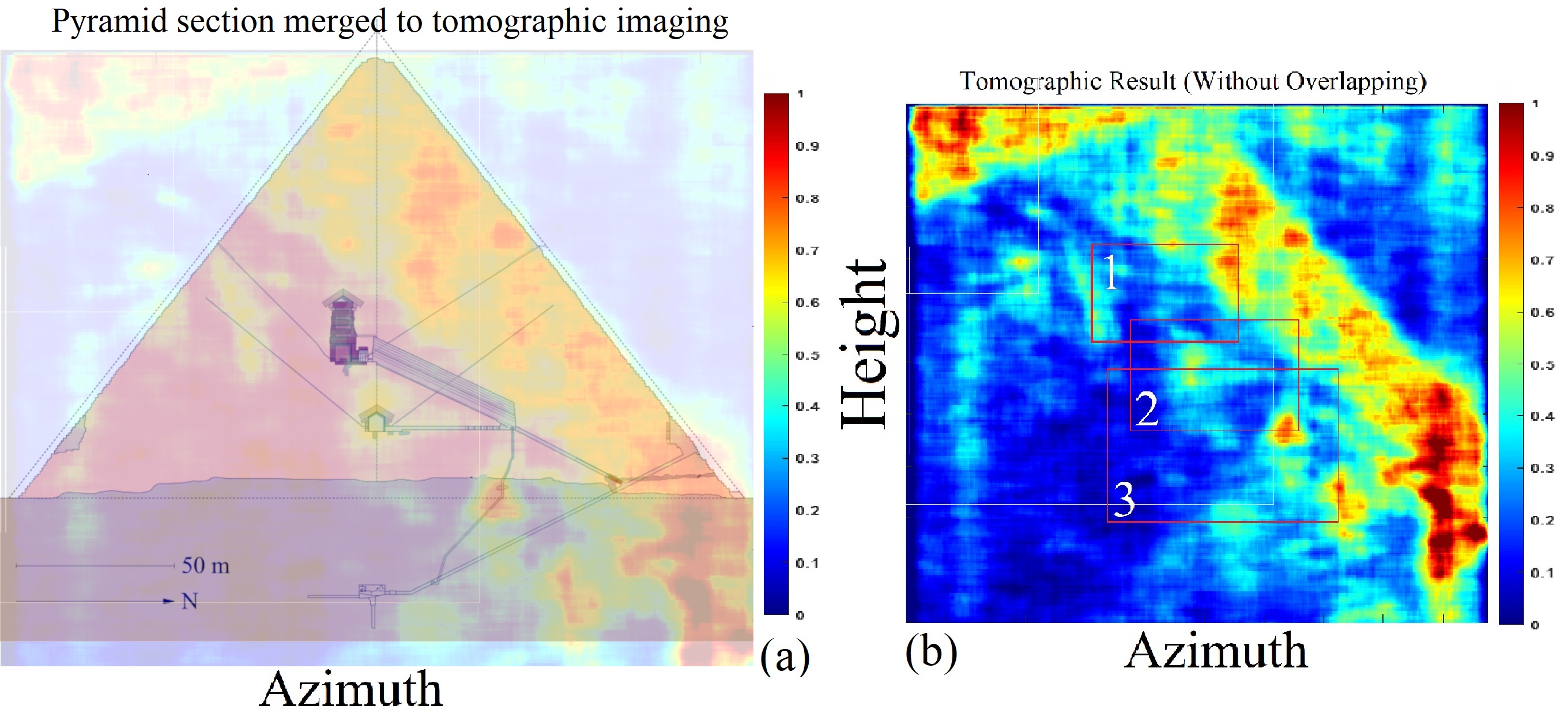}
	\caption{Schematic representation of the Khufu pyramid.}
	\label{Merged_1}
\end{figure}

\begin{figure} 
	\centering
	\includegraphics[width=15.0cm,height=4.5cm]{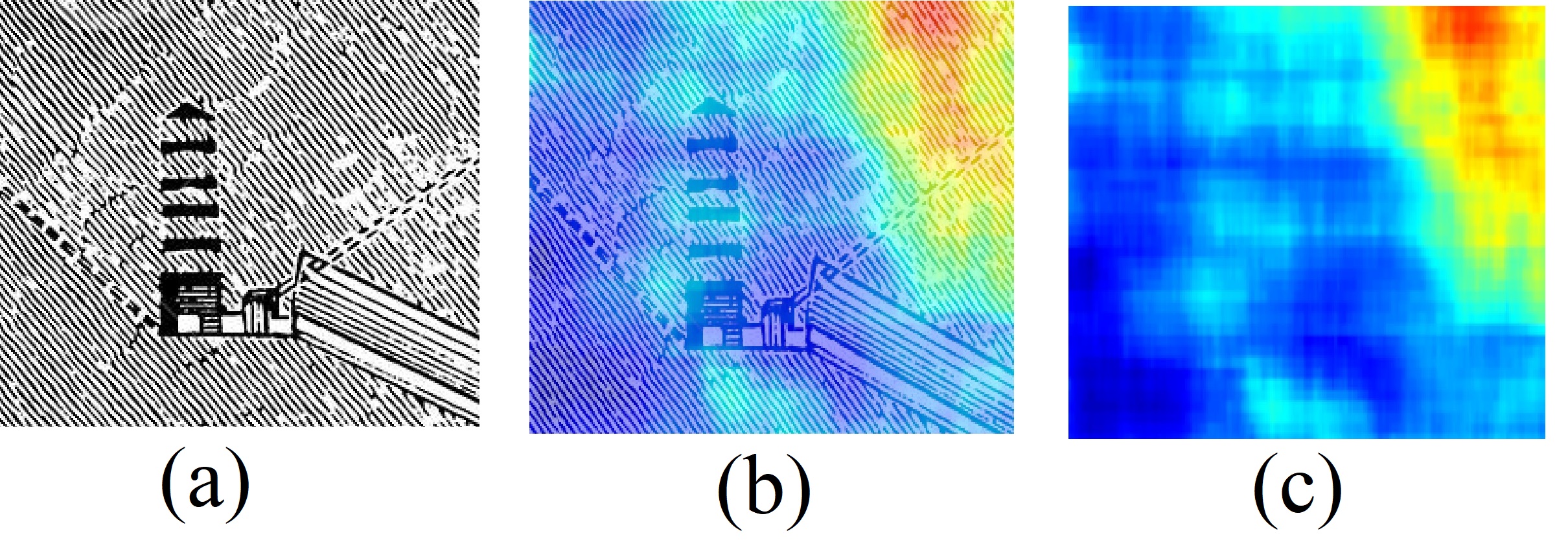}
	\caption{Imagine SAR of the pyramid of Khnum-Khufu. The V-shaped plot represents the tomographic line for which the vibrations were calculated, so that the inner section of the pyramid can be represented.}
	\label{SLC_3}
\end{figure}
\section{Internal Experimental Results of Khnum-Khufu}\label{Experimental_Rasults_Internal}

Data analysis obtained using te SAR tomographic Doppler imaging technique was able to provide clear objective elements to understand the internal structure of the pyramid of Khnum-Khufu. When proposing our results, we start by describing the well-known structures and then move on to the description of all the unknown structures.
The internal imaging obtained from multiple angles, allow us to obtain an accurate 3D model that gave us the possibility, like never before, to take a look inside one of the most important and mysterious megalithic monuments in the world.
This subsection describes all the tomographic measurements and the entire internal architecture of Khnum-Khufu has been redesigned, which we propose in Figure \ref{Fully_3D_1}. The complete list of rooms, corridors and tunnels that had never been inventoried until now is shown in Tab. \ref{Tab_1}. Each structure (simple or complex) was assigned to a unique tag numbered from 1 to 20, according to Figure \ref{Fully_3D_1}. For this work, we have not used any kind of simulated data or predictive mathematical model, but rather we report in a scientific manner what the CSG satellite has brought to our attention. Here we list the explanation of all the results obtained respecting the order given in Tab. \ref{Tab_1}. Each structure listed in Tab. \ref{Tab_1} will be described in detail, identified in the tomography and reconstructed within a Computer-Aided Design and Drafting (CAD) environment where all measurements will be provided.

\begin{figure} 
	\centering
	\includegraphics[width=15.0cm,height=3.5cm]{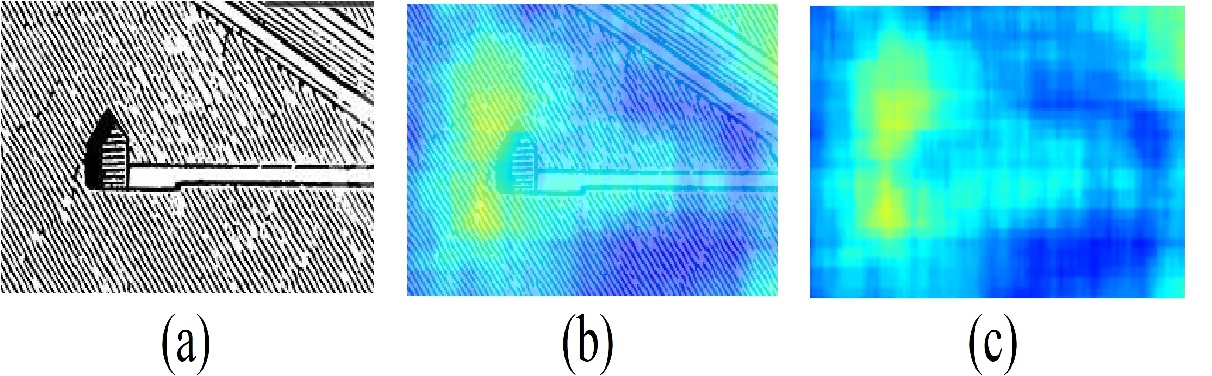}
	\caption{(a): Schematic representation of the ZED (box 1 details of Figure \ref{Merged_1} (b)). (b): (b): Schematic representation of the ZED (box 1 details of Figure \ref{Merged_1} (b)) partially overlapped to tomographic result. (c): Figure \ref{Merged_1} (b) box 1 details Tomographic result.}
	\label{SLC_4}
\end{figure}

\begin{figure} 
	\centering
	\includegraphics[width=15.0cm,height=6.0cm]{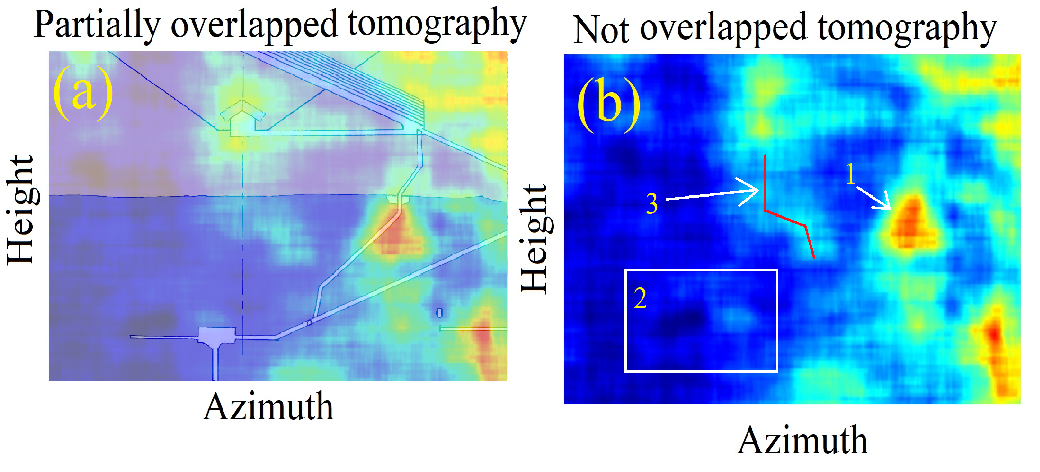}
	\caption{(a): Schematic representation of Figure \ref{Merged_1} (b) box 3 details partially overlapped with tomographic result. (b): tomographic result of Figure \ref{Merged_1} (b) box 3 details.}
	\label{Connessione_Reg_Sotto}
\end{figure}

\subsection{Imaging of known structures}
Before describing what has been discovered (what is still unknown) we propose the imaging of the known objects, in particular the King's room, the Zed, the Queen's room, the Grand Gallery, the grotto, and the so-called unfinished room. The SAR image of the pyramid is shown in Figure \ref{Merged_0} (a), while the internal diagram of the pyramid, oriented towards the North (the Northern direction goes from left to right) is shown in Figure \ref{Merged_0} (b). The first tomographic result is presented in Figure \ref{Merged_1} (a) and (b). This result is illustrated in Figure \ref{Merged_0} (a) by means of two yellow lines (identified by the number 1) that extend from the ground towards the pyramid apex. By estimating the vibrations, the pyramid is transparent like a crystal due to their penetration characteristics within the solid rock, and its internal structures can be observed in Figure {\ref{Merged_1} (a) and (b). Figure \ref{Merged_1} (a) represents the sonic tomography partially overlapped with the picture in Figure \ref{Merged_0} (b), while Figure \ref{Merged_0} (b) shows the non-overlapped tomography where three areas of interest are shown, and details are below studied. Figure \ref{SLC_3} (a), (b) and (c) is the detailed representation of the ZED, located inside box number 1 of Figure \ref{Merged_1}(b), where \ref{SLC_3} (a) is the schematic representation, while Figures \ref{SLC_3} (b) and (c) are the partially overlapped, and not-overlapped tomography magnitude images respectively. The Queen's chamber particular is depicted in Figure \ref{SLC_4} (a), (b) and (c), here we use the same representation strategy where \ref{SLC_4} (a) is the room scheme and Figures \ref{SLC_4} (b) and (c) are the partially overlapped and not-overlapped tomography respectively. Figure \ref{Connessione_Reg_Sotto} (a) and (b) is the detailed image contained within the red box 3 of Figure \ref{Merged_1} (b) and the void commonly referred to as 'the Grotto' is clearly visible. The lower chamber, the one commonly referred to as 'the unfinished room', is detected by radar, although with a weak signal, within the white box number 2. The Queen's room appears to be connected through a corridor connecting the grotto to the room below. This corridor, although it appears to be abruptly interrupted, is detected through a radar signature, which is marked by the white arrow number 3. It is assumed that this corridor follows the trajectory indicated by the red line, also pointed out by the white arrow number 3. Figure \ref{Knum_Khufu_1} (b) and (c) show the detailed figure of the ZED, the colossal monument located at the heart of the pyramid (visible in Figure \ref{Merged_1} (a)). Figure \ref{Knum_Khufu_1} (b) is its representation along the West-East direction, while Figure \ref{Knum_Khufu_1} (c) shows its pattern along the South-North direction. Concluding this section we propose in Figure \ref{ZED_12} (a) and (b) the partially and not-overlapped tomography of the Zed respectively, where also the King's chamber is visible.
	
Here we show in sequence all the tomographies we have calculated, whereby for each of them the subpicture (a) represents the SAR image in magnitude, whereby the tomographic line of investigation is visible (indicated with a yellow line marked with the number 1), while subpicture (b) represents the tomogram (again in magnitude). In order to be more clear, we report the aforementioned figures in Table \ref{Tab_2}:
\begin{figure} 
	\centering
	\includegraphics[width=15.0cm,height=7.0cm]{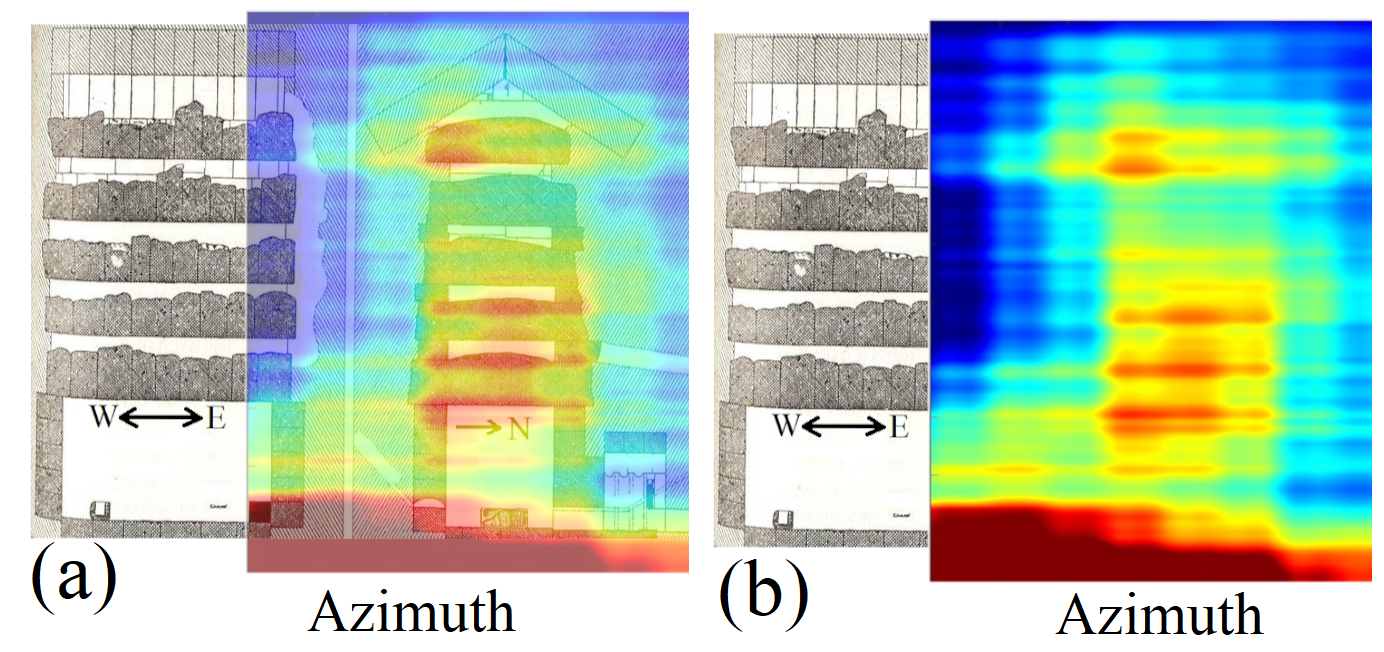}
	\caption{Schematic representation of the Zed and the King chamber.}
	\label{ZED_12}
\end{figure}
\begin{figure} 
	\centering
	\includegraphics[width=15.0cm,height=5.0cm]{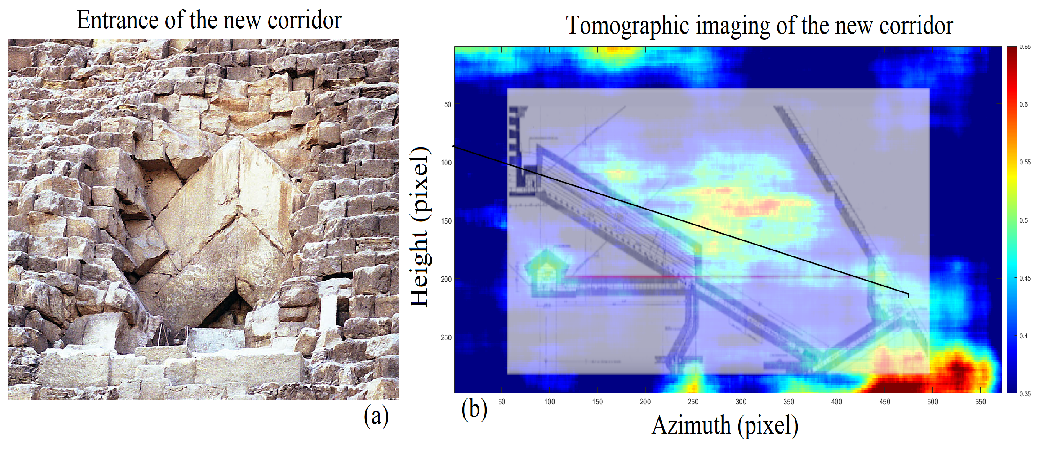}
	\caption{Entrance on Khnum-Khufu. (a): Optical image. (b): Tomographic map (magnitude).}
	\label{Paper_Ingresso_Principale_3}
\end{figure}
\begin{figure}[htp]
	\centering
	\includegraphics[width=15.0cm,height=6.0cm]{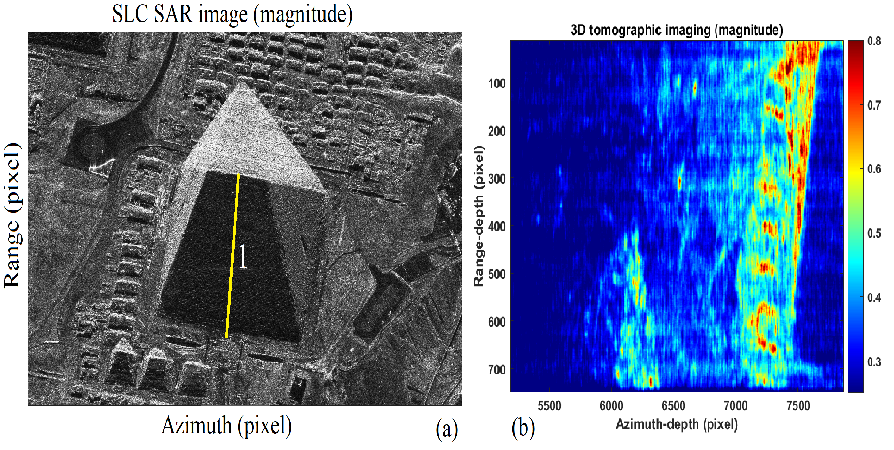}
	\caption{SAR images. (a): SLC SAR image (magnetude). (b): Tomographic result (magnetude).}
	\label{Mask_6}
\end{figure}

\begin{figure}[htp]
	\centering
	\includegraphics[width=15.0cm,height=6.0cm]{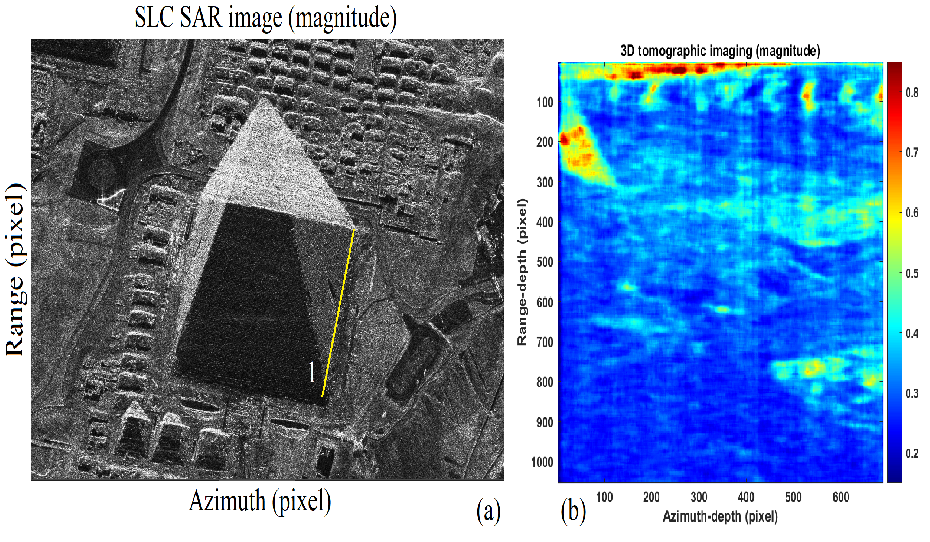}
	\caption{SAR images. (a): SLC SAR image (magnetude). (b): Tomographic result (magnetude).}
	\label{Mask_7}
\end{figure}

\begin{figure}[htp]
	\centering
	\includegraphics[width=15.0cm,height=6.0cm]{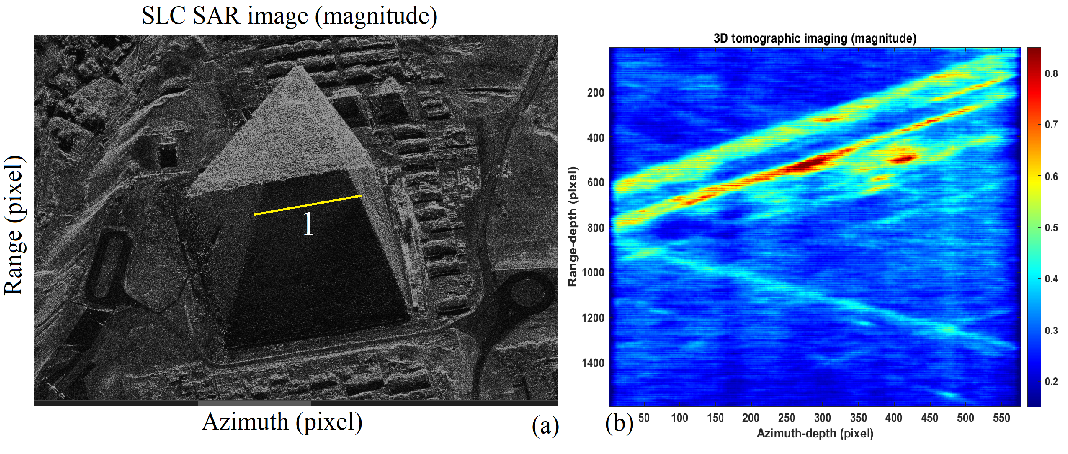}
	\caption{SAR images. (a): SLC SAR image (magnetude). (b): Tomographic result (magnetude).}
	\label{Mask_7_1}
\end{figure}

\begin{figure}[htp]
	\centering
	\includegraphics[width=15.0cm,height=6.0cm]{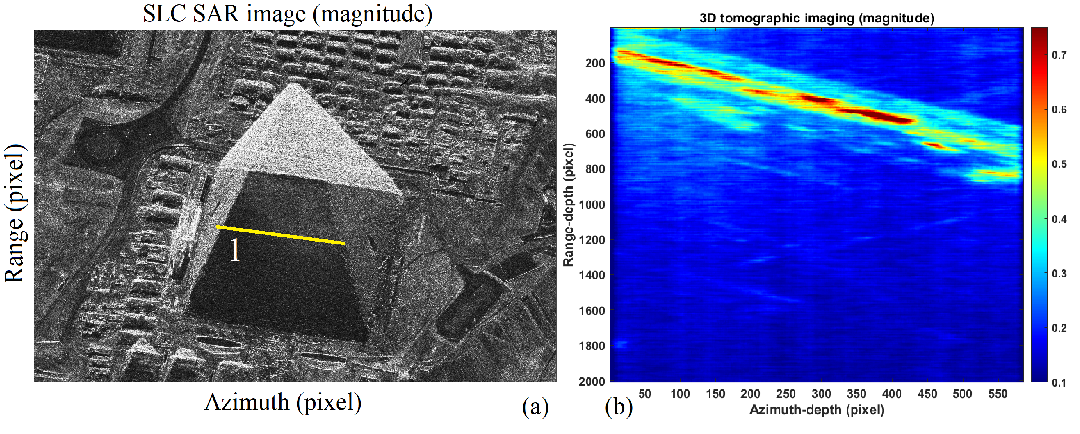}
	\caption{SAR images. (a): SLC SAR image (magnetude). (b): Tomographic result (magnetude).}
	\label{SLC_2_1}
\end{figure}

\begin{figure}[htp]
	\centering
	\includegraphics[width=15.0cm,height=6.0cm]{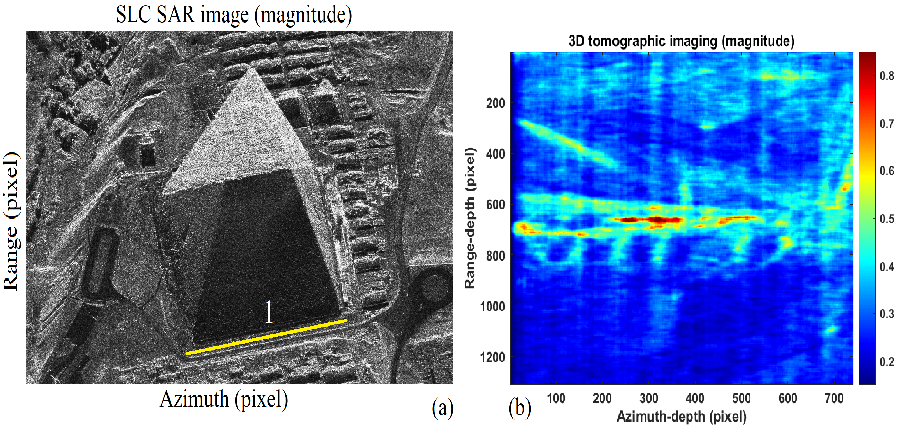}
	\caption{SAR images. (a): SLC SAR image (magnetude). (b): Tomographic result (magnetude).}
	\label{Linea_Tomografica_3}
\end{figure}

\begin{figure}[htp]
	\centering
	\includegraphics[width=15.0cm,height=6.0cm]{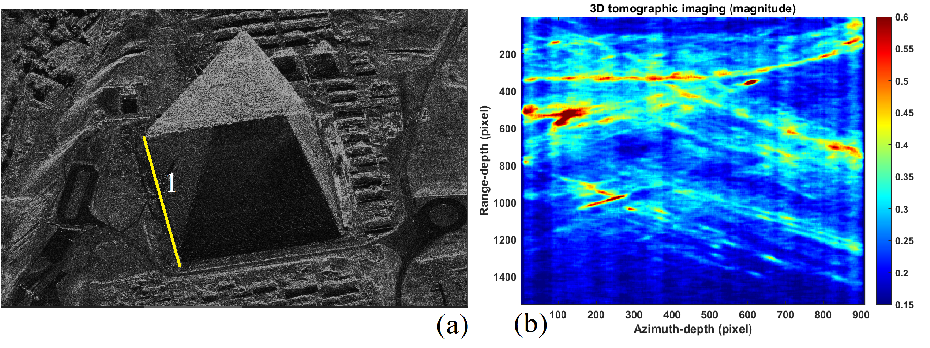}
	\caption{SAR images. (a): SLC SAR image (magnetude). (b): Tomographic result (magnetude).}
	\label{Linea_Tomografica_2}
\end{figure}

\begin{figure}[htp]
	\centering
	\includegraphics[width=15.0cm,height=6.0cm]{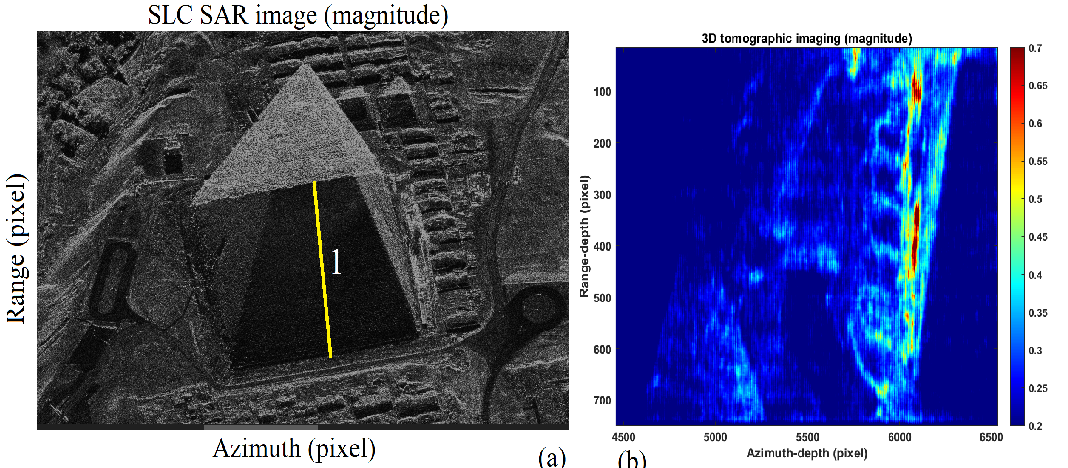}
	\caption{SAR images. (a): SLC SAR image (magnetude). (b): Tomographic result (magnetude).}
	\label{Linea_Tomografica_2_1}
\end{figure}

\begin{figure}[htp]
	\centering
	\includegraphics[width=15.0cm,height=6.0cm]{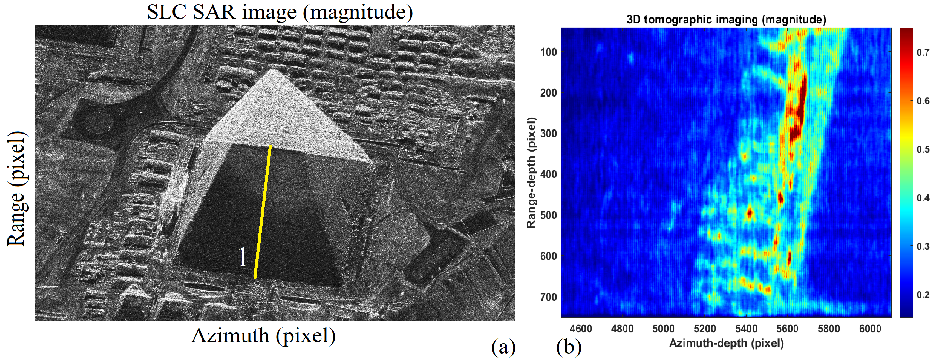}
	\caption{SAR images. (a): SLC SAR image (magnetude). (b): Tomographic result (magnetude).}
	\label{SLC_3_1}
\end{figure}

\begin{figure}[htp]
	\centering
	\includegraphics[width=15.0cm,height=6.0cm]{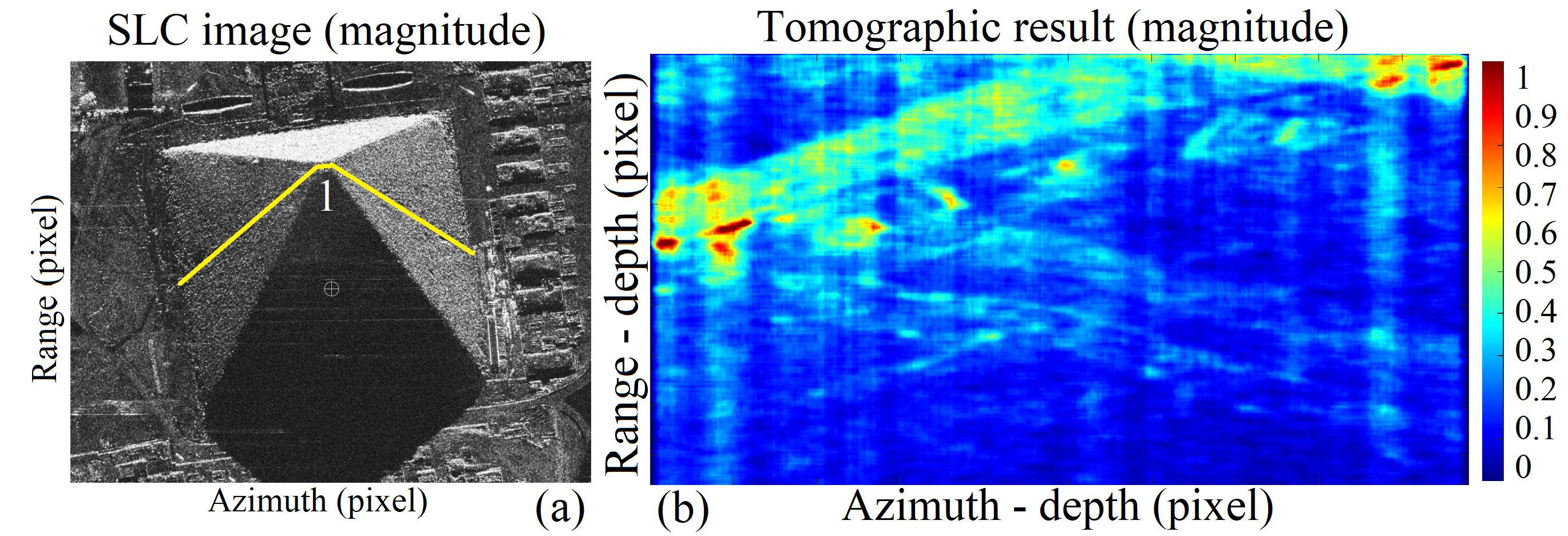}
	\caption{SAR images. (a): SLC SAR image (magnetude). (b): Tomographic result (magnetude).}
	\label{Tomo_1_1}
\end{figure}

\begin{figure}[htp]
	\centering
	\includegraphics[width=15.0cm,height=6.0cm]{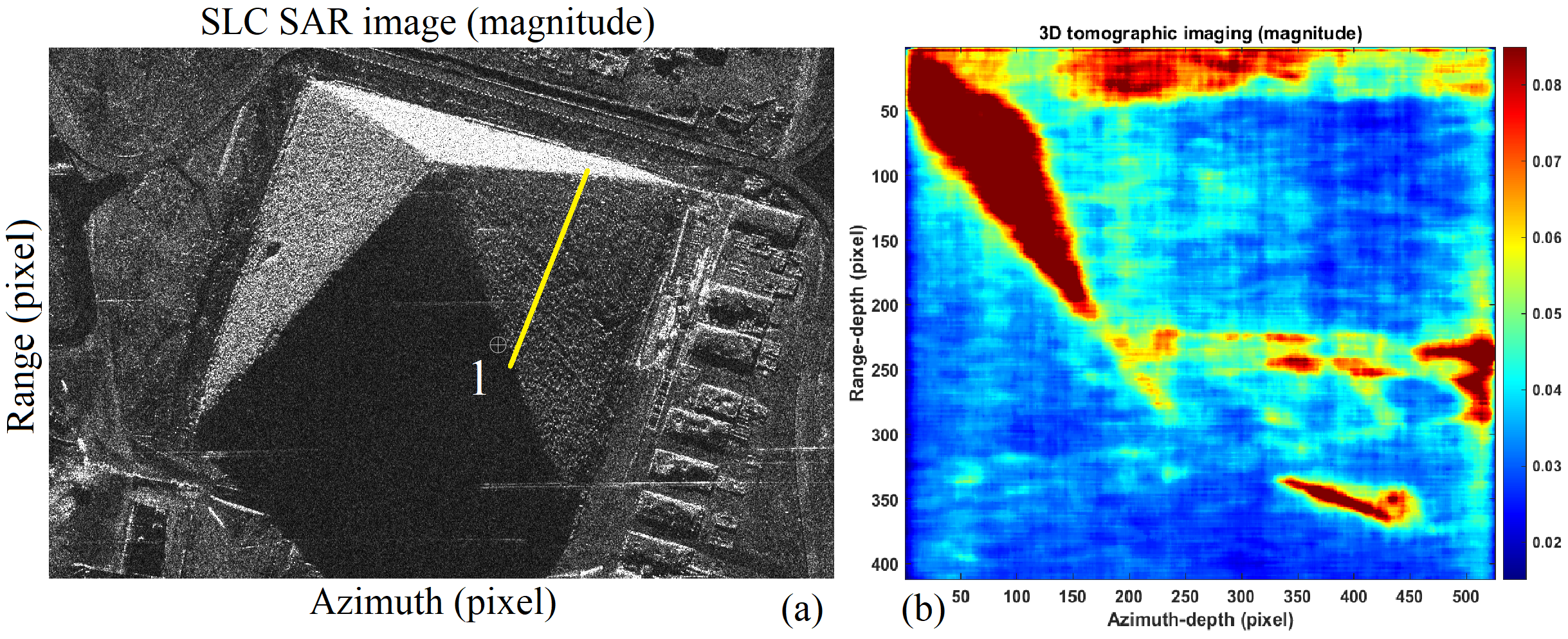}
	\caption{SAR images. (a): SLC SAR image (magnetude). (b): Tomographic result (magnetude).}
	\label{Tomo_1_2}
\end{figure}

\subsection{Eastern and Western ascending ramps (tag 1, tag 2)}
Two inclined and diverging ramps (identified with the numbers 1 and 2 in the 3D reconstruction depicted in Figure \ref{Fully_3D_1}), characterized by an approximate slope of about 42 degrees, located inside the West and East sides. For both ramps, the lower part starts from the ground level on the North side and reaches half the height of the pyramid on the south side. The reference images are Figure \ref{Tag_1_4_7} (a), (b), for the Eastern side, and Figure \ref{Tag_1_9} (a), (b), for the Western side, where the 3D models are compared to measured tomograms. From the figures, tags number 1,2,3,4 5,7 and 9 are recognized. 
\begin{figure} 
	\centering
	\includegraphics[width=15.0cm,height=6.0cm]{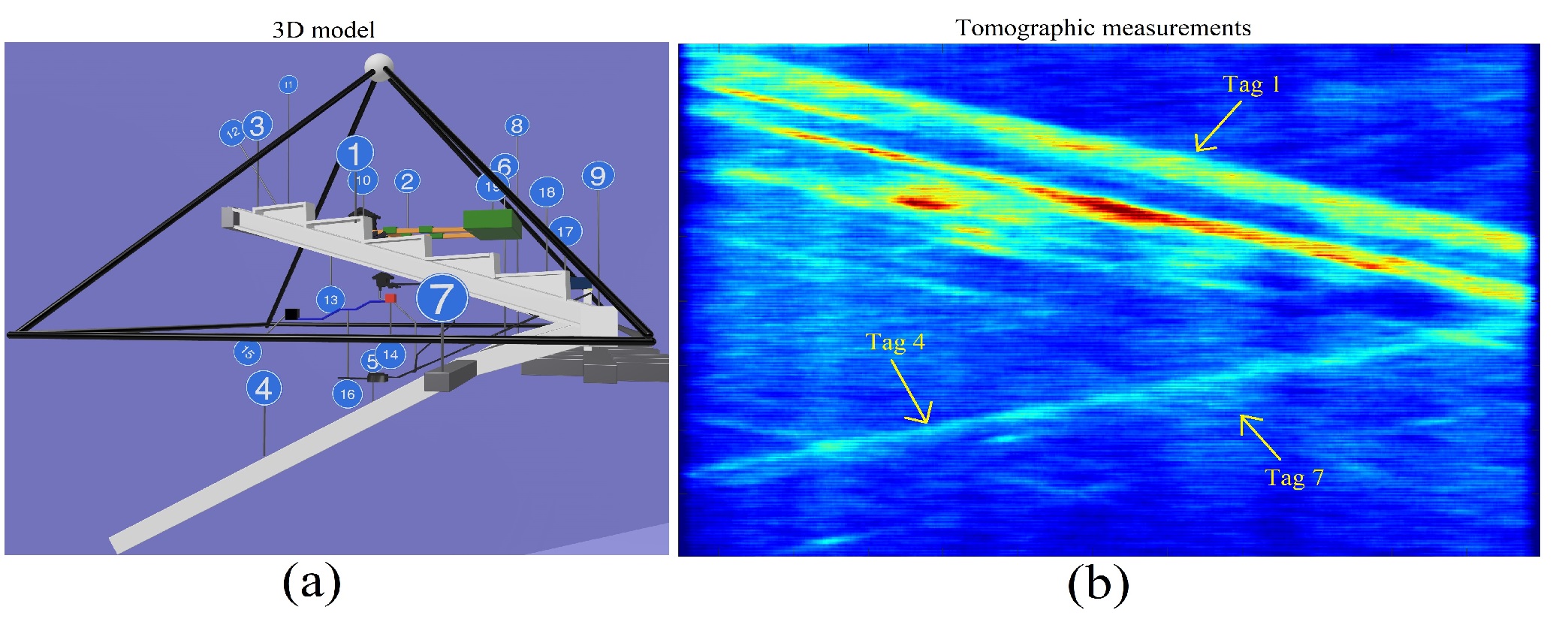}
	\caption{Tags association from tomography to 3D model. (a): 3D model of Khnum-Khufu. (b): Tomographic reconstruction (magnitude).}
	\label{Tag_1_4_7}
\end{figure}
\begin{figure} 
	\centering
	\includegraphics[width=15.0cm,height=6.0cm]{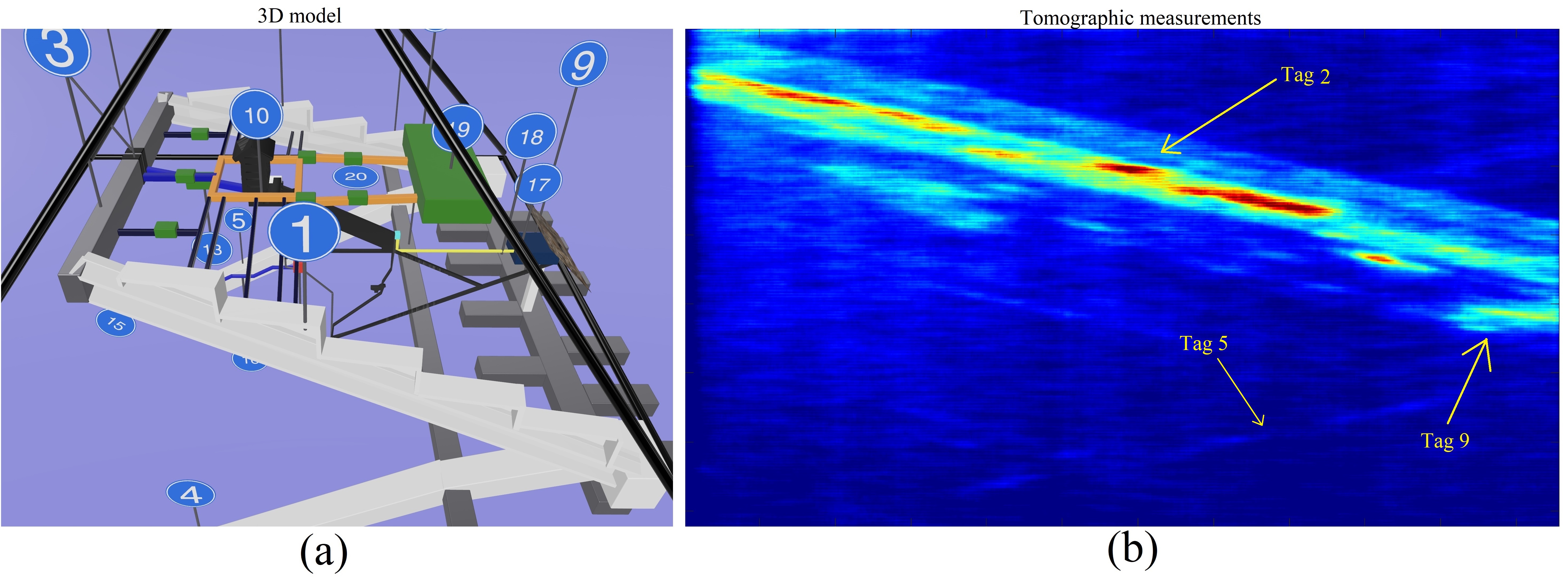}
	\caption{Tags association from tomography to 3D model. (a): 3D model of Khnum-Khufu. (b): Tomographic reconstruction (magnitude).}
	\label{Tag_1_9}
\end{figure}

\subsection{Southern Corridor, (tag 3)}
The ascending corridors are connected to each other by means of a horizontal structure placed at a height of about 90 meters and located near the south side of the pyramid (identified with the number 3 in Figure \ref{Fully_3D_1}). The corridor is recognized in Figure \ref{Tags_Generic_4} (b), where the 3D reconstruction can be seen in Figure \ref{Tags_Generic_4} (b).
\begin{figure} 
	\centering
	\includegraphics[width=15.0cm,height=6.0cm]{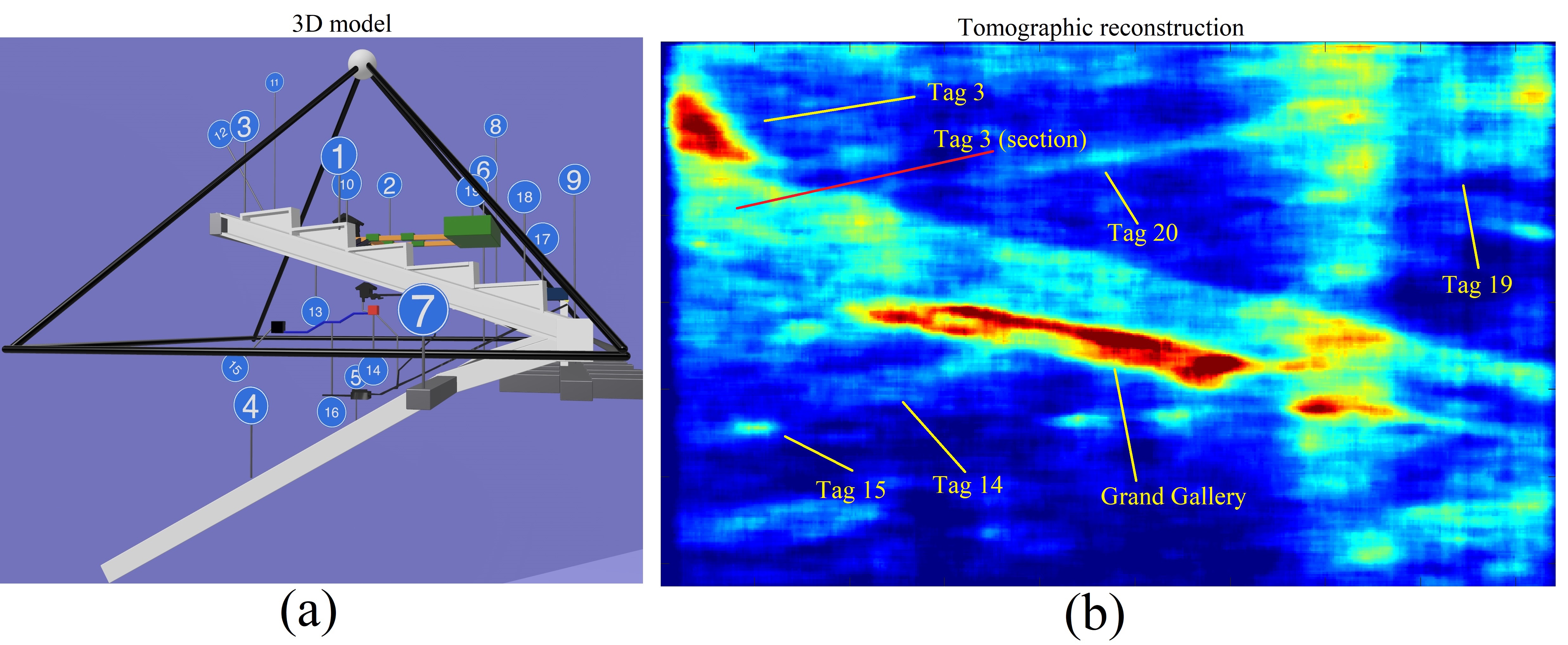}
	\caption{Tags association from tomography to 3D model. (a): 3D model of Khnum-Khufu. (b): Tomographic reconstruction (magnitude).}
	\label{Tags_Generic_4}
\end{figure}

\subsection{Eastern and Western descending ramps, (tag 4, tag 5)}
Two ramps that, both connected to the previous ones, parallel to each other and also to the East and West base sides that run through a descending underground section with variable slope (numbers 4 and 5 in the 3D model). Figures \ref{Tag_4_5} (a) and \ref{Tag_4_5_2} (a) are the 3D reconstruction models showing the descending corridors from two different view angles, while Figures \ref{Tag_4_5} (b) and \ref{Tag_4_5_2} (b) corresponds to tomographic measurements of the same tags, corresponding to the same descending corridors.
\begin{figure} 
	\centering
	\includegraphics[width=15.0cm,height=6.0cm]{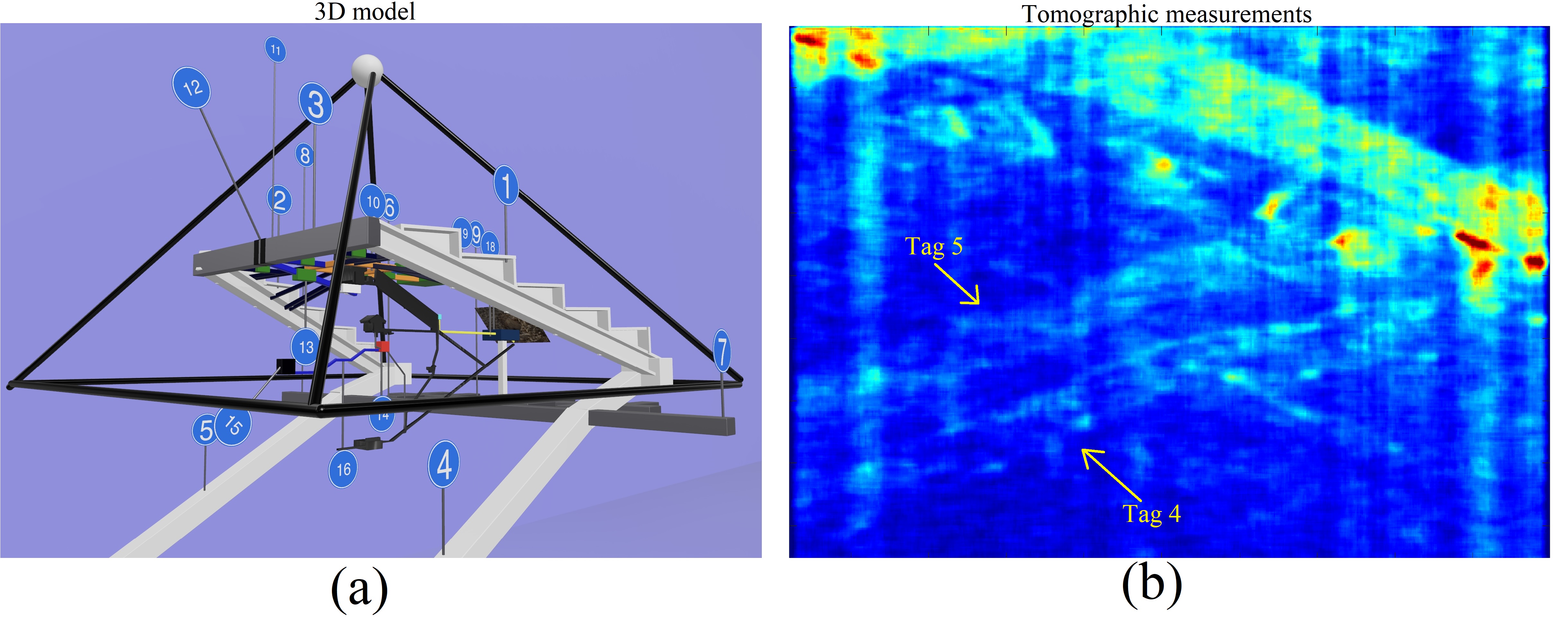}
	\caption{Tags association from tomography to 3D model. (a): 3D model of Khnum-Khufu. (b): Tomographic reconstruction (magnitude).}
	\label{Tag_4_5}
\end{figure}

\begin{figure} 
	\centering
	\includegraphics[width=15.0cm,height=6.0cm]{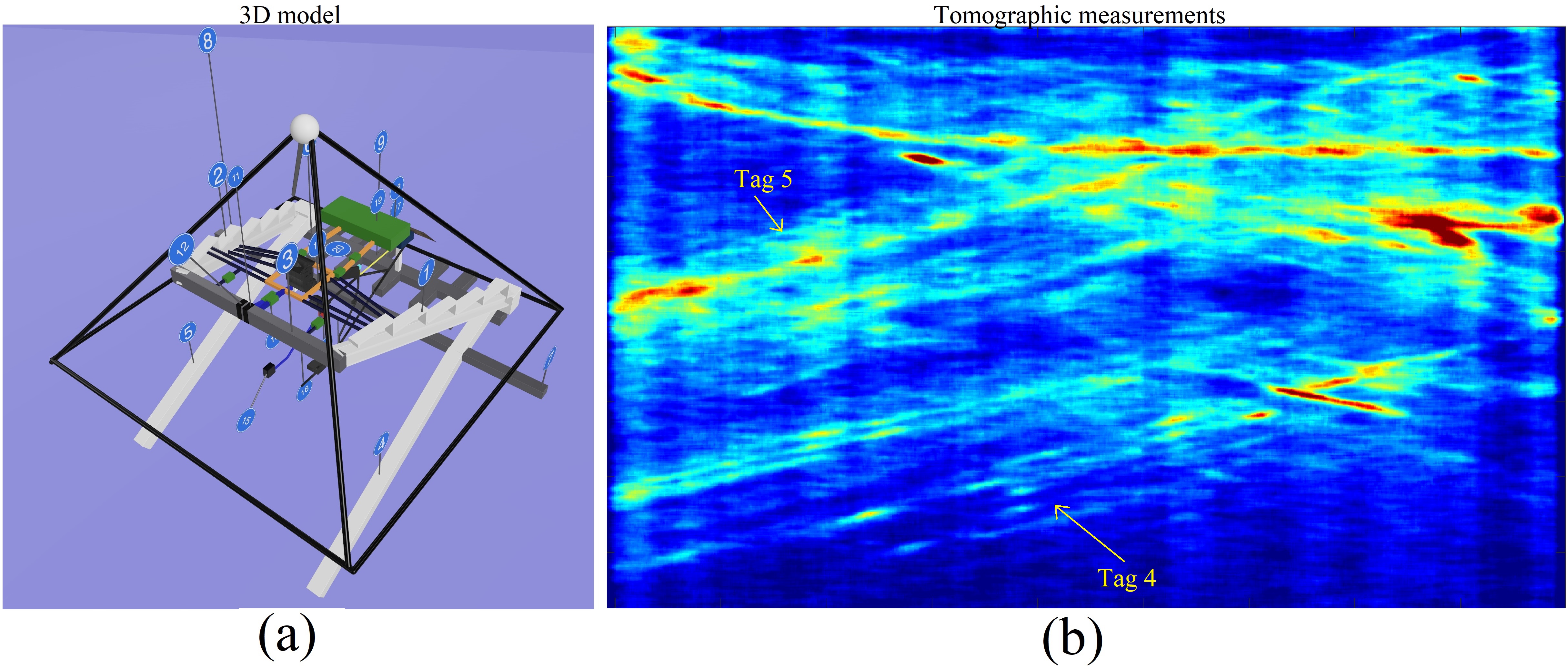}
	\caption{Tags association from tomography to 3D model. (a): 3D model of Khnum-Khufu. (b): Tomographic reconstruction (magnitude).}
	\label{Tag_4_5_2}
\end{figure}

\subsection{Northern underground corridor, (tag 6) and Northern-East and Northern-West underground corridors (tag 7, tag 8)}
At the point where descending ramps (tags 4 and 5) increase slope they appear connected with the Northern underground corridor (tag number 6 structure), parallel to the North side of the pyramid. The Northern underground corridor is characterized by two extrusions, which are still located underground. Figure \ref{Tag_6_7_8} (a) is the 3D reconstruction model showing the structures tagged by numbers 6, 7 and 8, while \ref{Tag_6_7_8} (b) corresponds to tomographic measurements of the same tags, corresponding to the same descending corridors. The section of this structure is deducted on the tomogram of Figure \ref{Tags_Generic_2} (b), while the corresponding 3D model is showed in Figure \ref{Tags_Generic_2} (a).

\begin{figure} 
	\centering
	\includegraphics[width=15.0cm,height=6.0cm]{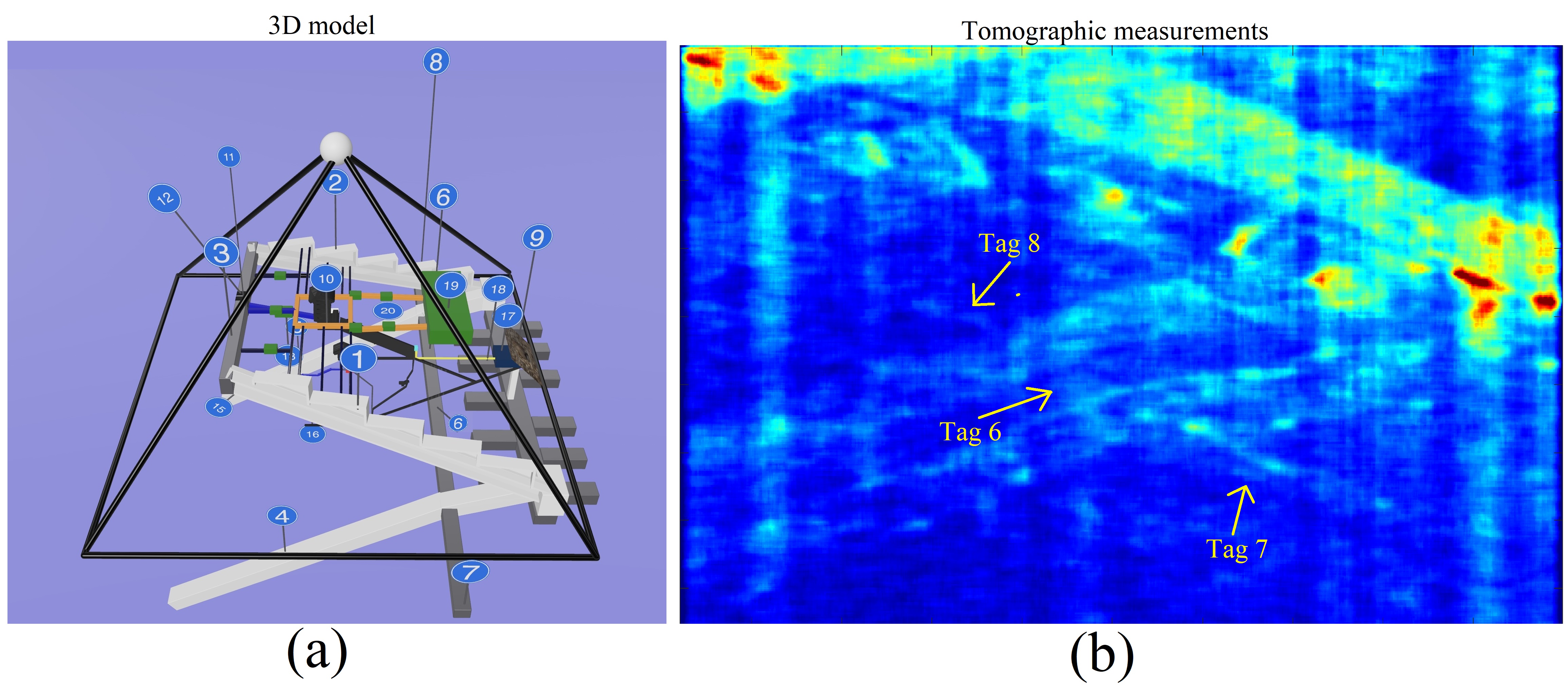}
	\caption{Tags association from tomography to 3D model. (a): 3D model of Khnum-Khufu. (b): Tomographic reconstruction (magnitude).}
	\label{Tag_6_7_8}
\end{figure}

\begin{figure} 
	\centering
	\includegraphics[width=15.0cm,height=6.0cm]{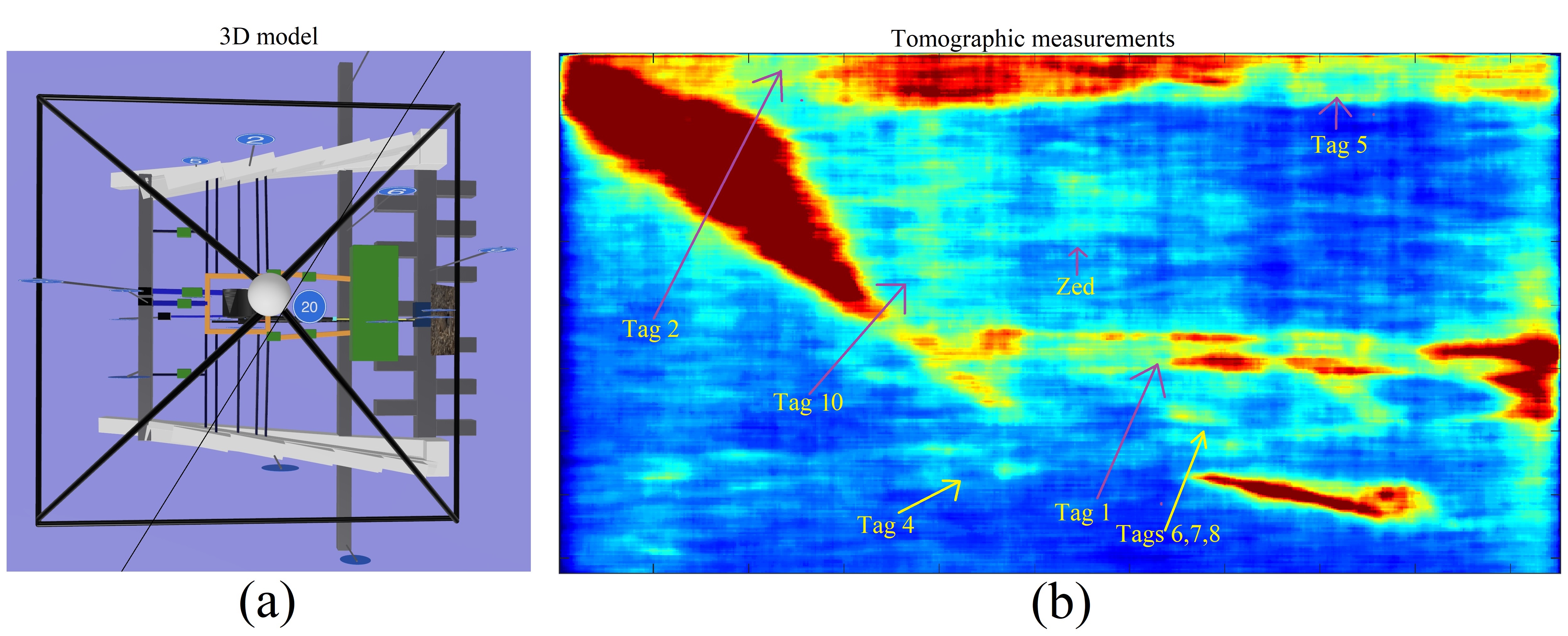}
	\caption{Tags association from tomography to 3D model. (a): 3D model of Khnum-Khufu. (b): Tomographic reconstruction (magnitude).}
	\label{Tags_Generic_2}
\end{figure}
\begin{figure} 
	\centering
	\includegraphics[width=16.0cm,height=4.0cm]{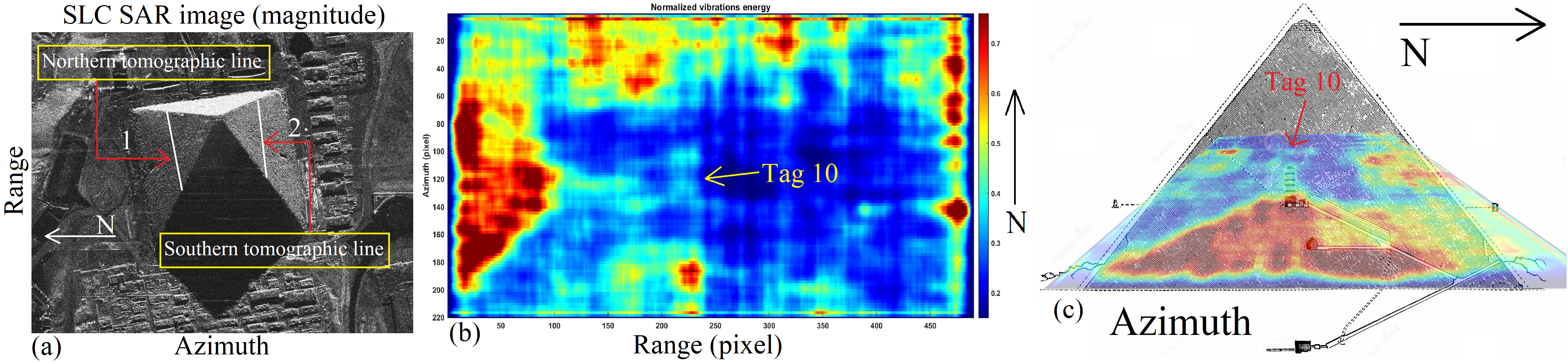}
	\caption{SLC SAR image (magnitude).}
	\label{Tomografia_Orizzontale_1}
\end{figure}
\begin{figure} 
	\centering
	\includegraphics[width=15.0cm,height=6.0cm]{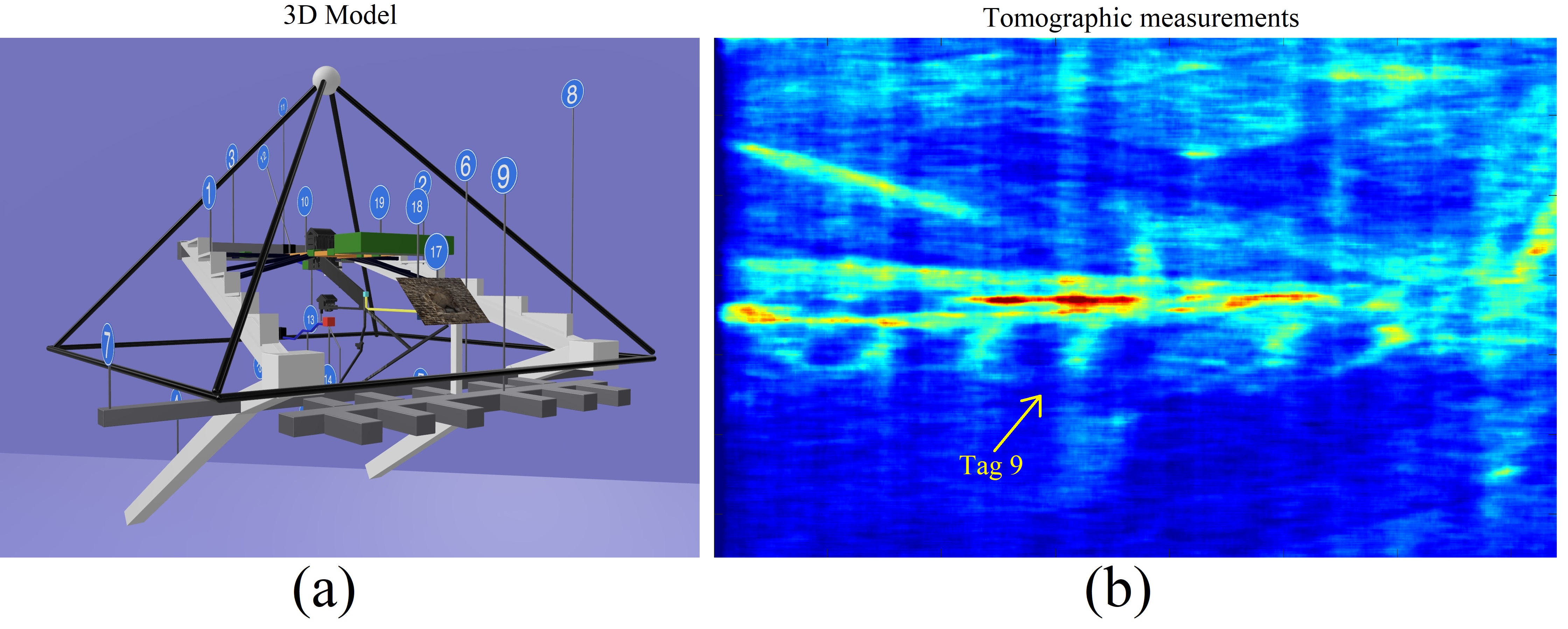}
	\caption{Tags association from tomography to 3D model. (a): 3D model of Khnum-Khufu. (b): Tomographic reconstruction (magnitude).}
	\label{Tag_9}
\end{figure}
\begin{figure} 
	\centering
	\includegraphics[width=15.0cm,height=6.0cm]{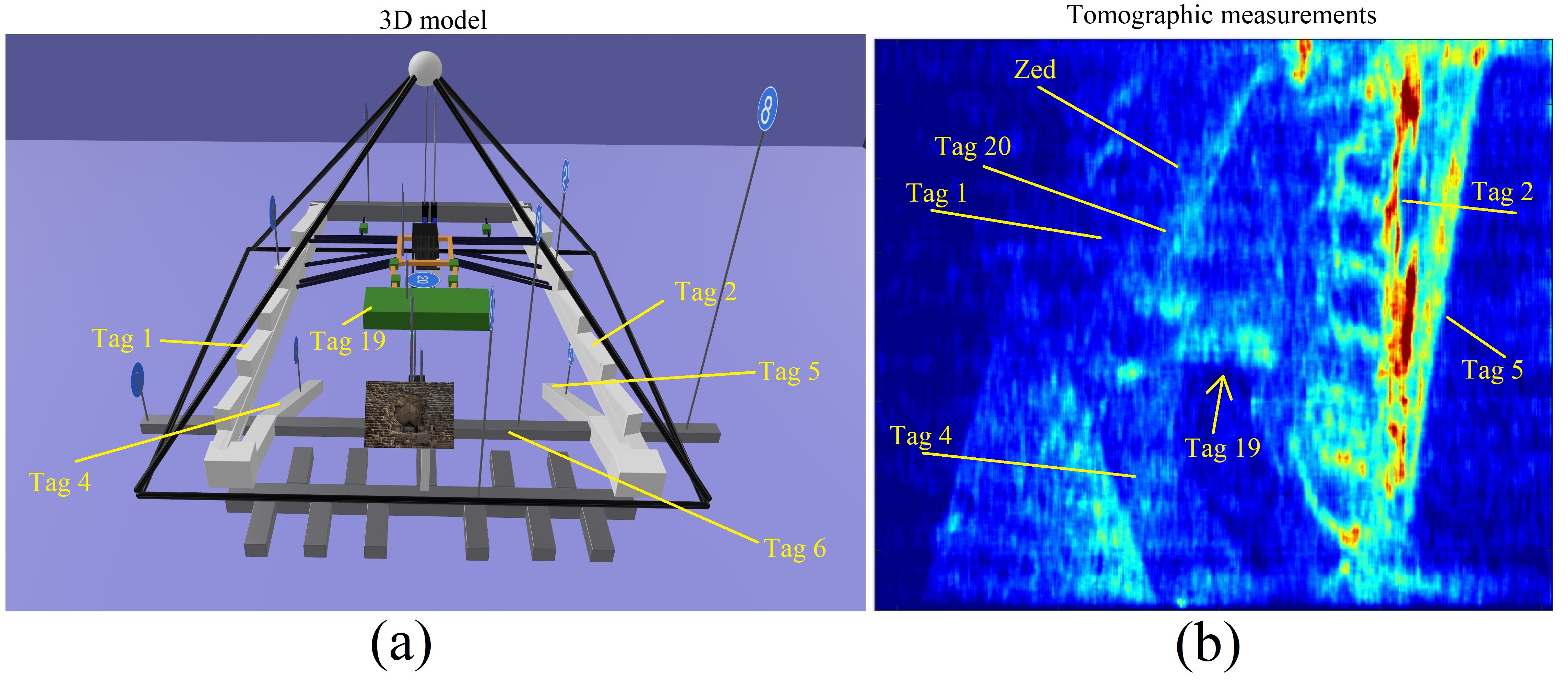}
	\caption{Tags association from tomography to 3D model. (a): 3D model of Khnum-Khufu. (b): Tomographic reconstruction (magnitude).}
	\label{Tags_Generic_1}
\end{figure}
\subsection{Northern underground complex-structure (tag 9)}
Immediately below the base of the pyramid structure, at the North side, a complex structure appears consisting of a horizontal body from which several identical bodies branch off, extruded perpendicularly to the main structure and characterized by a geometry, also present in other Egyptian pyramids, such as the pyramid of Zawyet El-Aryan \cite{dodson2000layer,noc2015analyse,bergendorff2019social} and the Sekhemkhet pyramid \cite{baud2015djeser} . This complex structure (Number 9 in the 3D model) is characterized by a small conduit placed in a central position that runs a short distance in a vertical direction, in analogy with the presence of a similar building also in El-Aryan \cite{dodson2000layer,noc2015analyse,bergendorff2019social} and Saqqara \cite{baud2015djeser}. The reference tomography is shown in Figure \ref{Tag_9} (b), while the 3D model is depicted in Figure \ref{Tag_9} (a). 

\subsection{ZED complex-structure (tag 10)}
A complex square structure (identified with the number 10 tag), which connects itself to the structure number 11, belonging to the structure of passage number 3. The structure 10 develops around the Zed, approximately at the height of the lowest room (Davison's Chamber) \cite{hancock2011fingerprints}. The reference tomography is shown in Figure \ref{Tags_Generic_2} (b), while the 3D model is depicted in Figure \ref{Tags_Generic_2} (a). The structure is also detected through different tomograms depicted in Figure \ref{Tomografia_Orizzontale_1} (b), (c), where the reference SLC image is showed in Figure \ref{Tomografia_Orizzontale_1} (a), and the tomographic lines 1 and 2 are showed on the northern and southern pyramid surfaces respectively.

\begin{figure} 
	\centering
	\includegraphics[width=15.0cm,height=6.0cm]{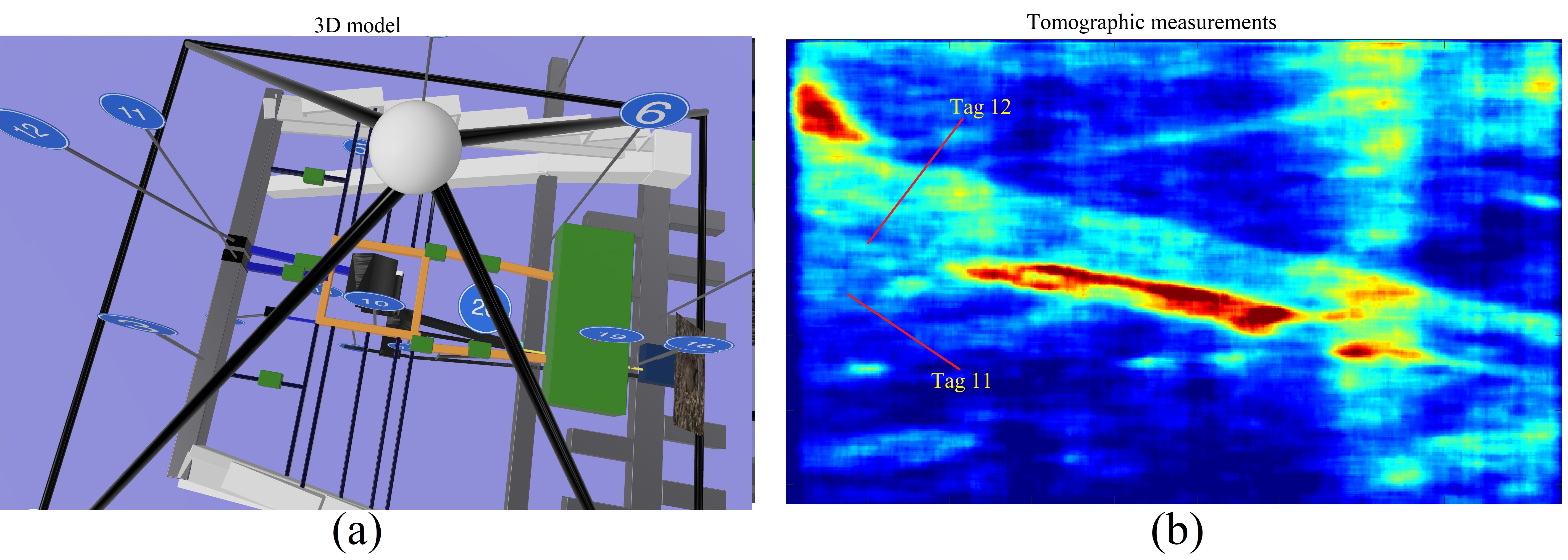}
	\caption{Tags association from tomography to 3D model. (a): 3D model of Khnum-Khufu. (b): Tomographic reconstruction (magnitude).}
	\label{Tags_Generic_5}
\end{figure}
\begin{figure} 
	\centering
	\includegraphics[width=15.0cm,height=6.0cm]{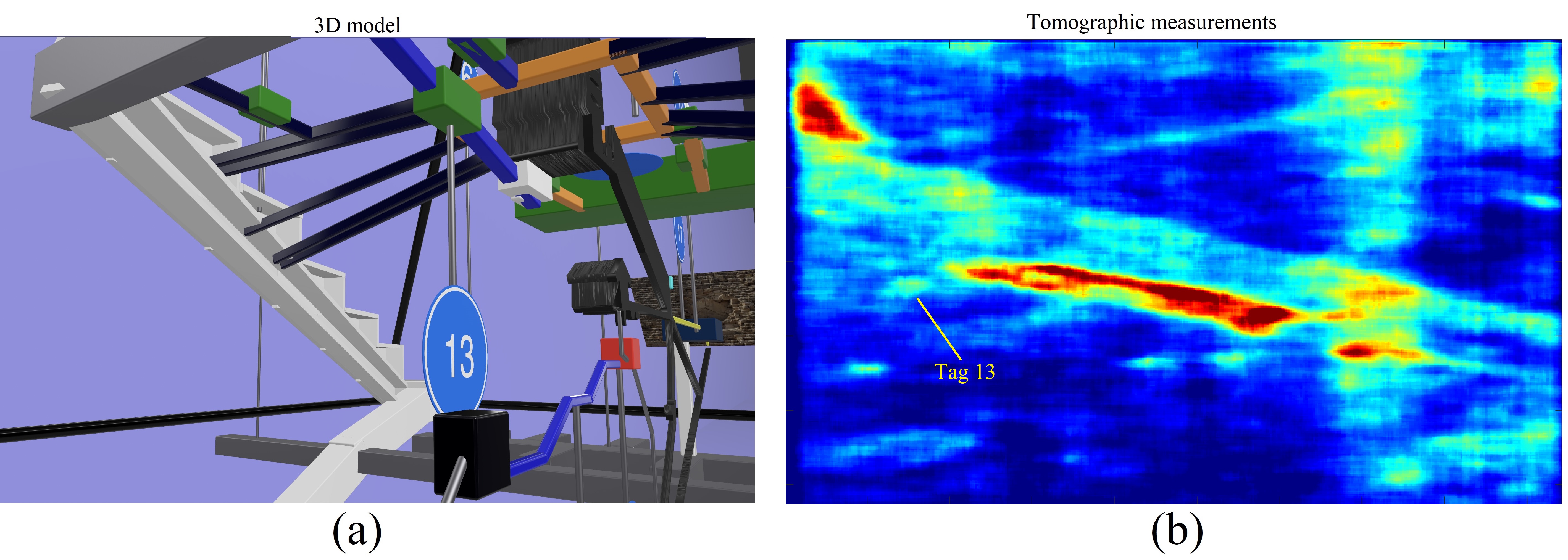}
	\caption{Tags association from tomography to 3D model. (a): 3D model of Khnum-Khufu. (b): Tomographic reconstruction (magnitude).}
	\label{Tags_Generic_6}
\end{figure}
\begin{figure} 
	\centering
	\includegraphics[width=15.0cm,height=6.0cm]{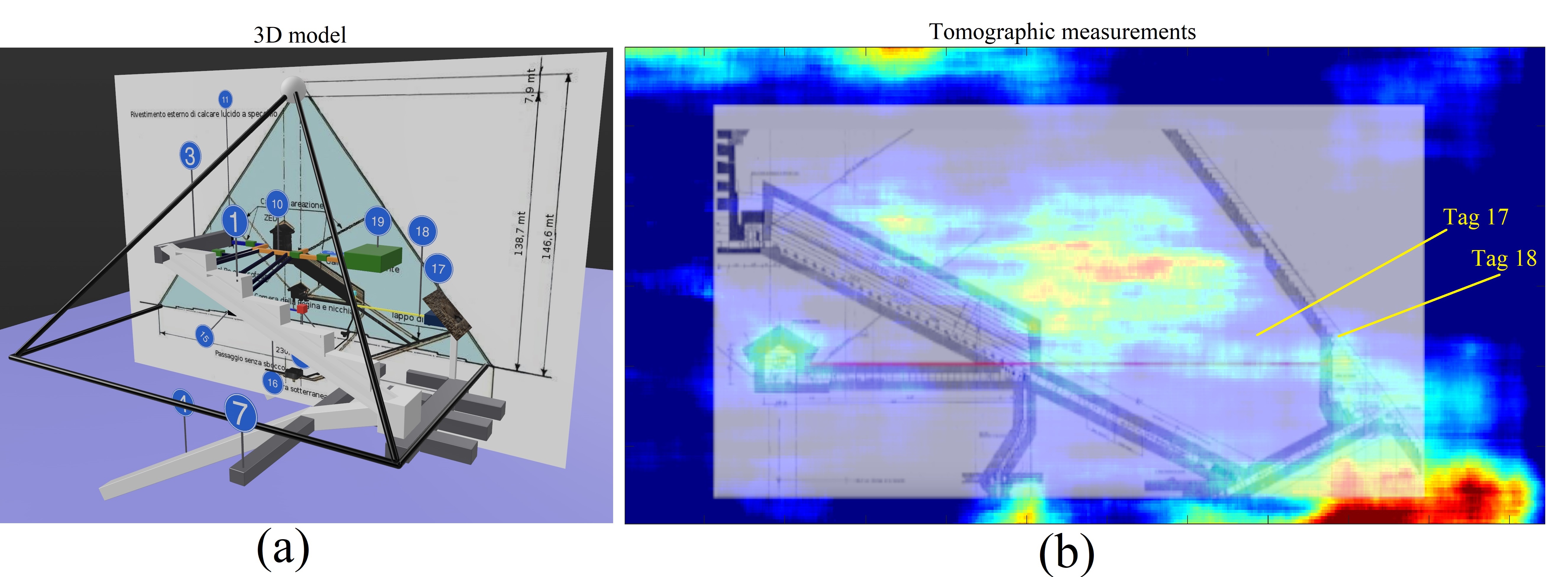}
	\caption{Tags association from tomography to 3D model. (a): 3D model of Khnum-Khufu. (b): Tomographic reconstruction (magnitude).}
	\label{Tag_Front_Corrider}
\end{figure}
\begin{figure} 
	\centering
	\includegraphics[width=15.0cm,height=6.0cm]{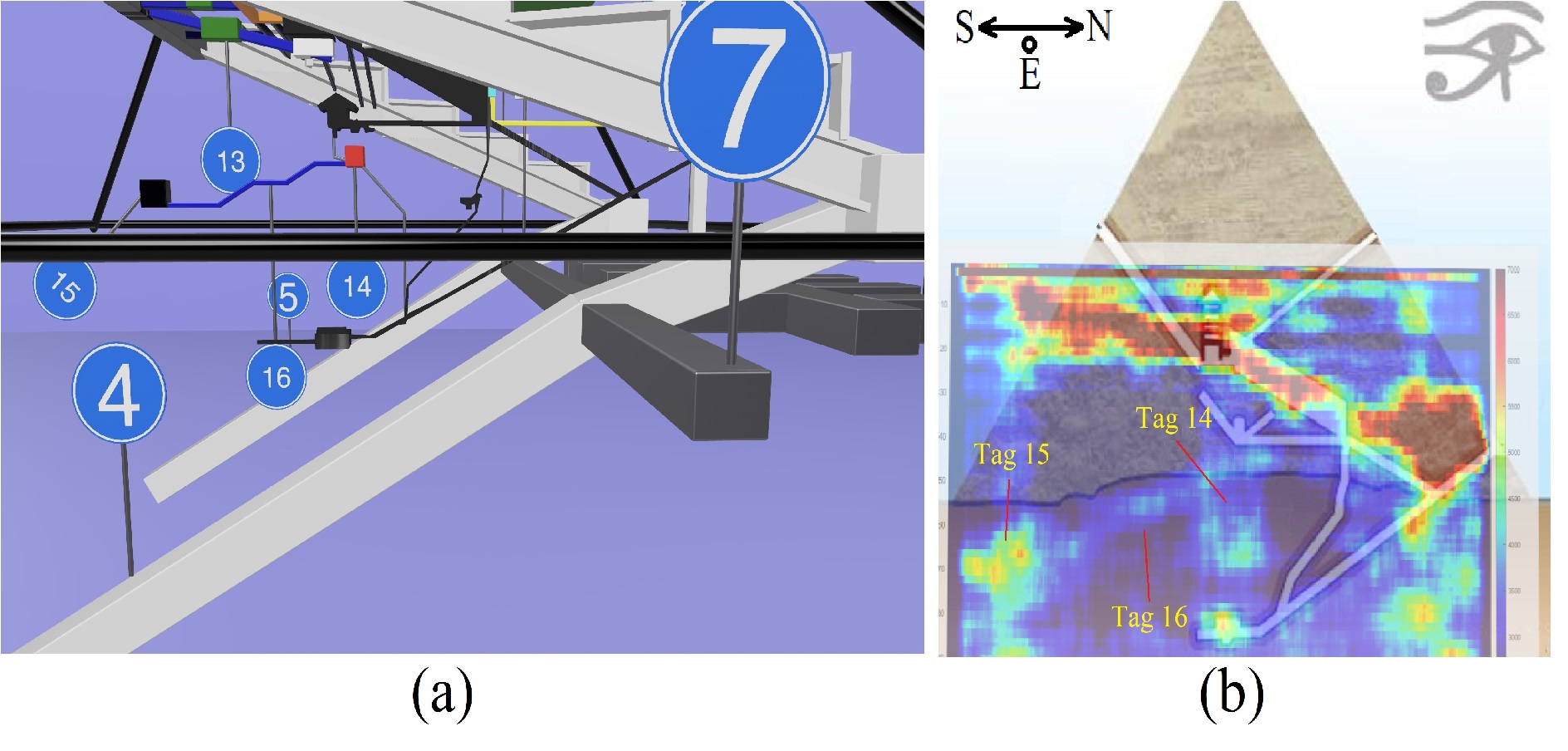}
	\caption{Tags association from tomography to 3D model. (a): 3D model of Khnum-Khufu. (b): Tomographic reconstruction (magnitude).}
	\label{IMG_2065}
\end{figure}

\subsection{Eastern and western sarcophagus passages facility (tag 11 and tag 12)}
Structure number 3 also seems to contain two sub-structures, identified with the numbers 11 and 12, connected, through corridor 13, to the King's room, through a passage that seems located under the floor of the latter. The reference tomography is shown in Figure \ref{Tags_Generic_5} (b), while the 3D model is depicted in Figure \ref{Tags_Generic_5} (a).

\subsection{Bottom sarcophagus room facility (tag 13)}
A room located below structures 11 and 12, connecting facilities 3 to 13. The reference tomography is shown in Figure \ref{Tags_Generic_6} (b), while the 3D model is depicted in Figure \ref{Tags_Generic_6} (a). The reference tomography is shown in Figure \ref{IMG_2065} (b), while the 3D model is depicted in Figure \ref{IMG_2065} (a).

\subsection{Queen's bottom room (tag 14)}
A further structure (identified with the number 14 in the 3D model), located below the Queen's chamber and connected to it through a small conduit. From space number 14 the conduit seems to continue, bonding a similar path which, from the room known as "Grotto", leads to the underground room called unfinished". The reference tomography is shown in Figure \ref{IMG_2065} (b), while the 3D model is depicted in Figure \ref{IMG_2065} (a).

\subsection{Southern bottom room (tag 15)}
A room located on the bottom of structures 11 and 12. The reference tomography is shown in Figure \ref{IMG_2065} (b), while the 3D model is depicted in Figure \ref{IMG_2065} (a).

\subsection{Southern Connection (tag 16)}
A further conduit (number 16 of the 3D model) joins the structure 14 to a structure, placed almost at ground level (number 15 of the 3D model). This facility has been discovered through tomographic result depicted in Figure \ref{IMG_2065} (b), where a particular of the 3D model is visible in Figure \ref{IMG_2065} (a).

\subsection{Little-void (tag 17)}
A void that can be located immediately behind the original entrance of the pyramid, not easily identificable in shape and size (number 17 of the 3D model), from which a horizontal conduit (number 18 of the 3D model) starts and which seems to end at the foot of the Grand Gallery, but not directly connected to it.
This is a void located in front of the Northern entrance of the pyramid \cite{morishima2017discovery}. The room is clearly visible in Figure \ref{Paper_Ingresso_Principale_3} (b), precisely located above the corridor identified by the structure with tag 18. It possible to observe the little-void also on Figure \ref{Tag_Front_Little-Void} (b), there the reference 2D model is reported in Figure \ref{Tag_Front_Little-Void} (a). 

\begin{figure} 
	\centering
	\includegraphics[width=15.0cm,height=6.0cm]{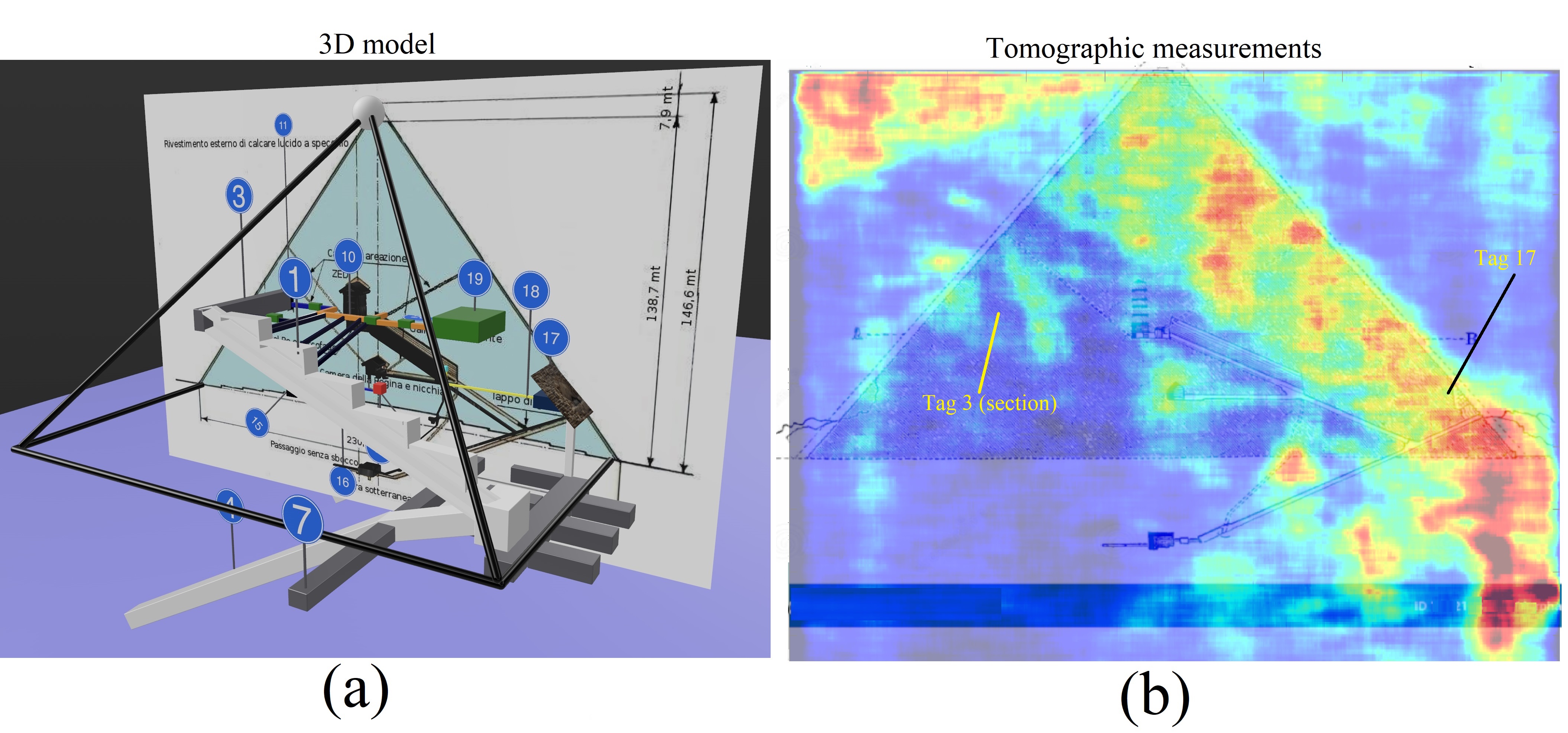}
	\caption{Tags association from tomography to 3D model. (a): 3D model of Khnum-Khufu. (b): Tomographic reconstruction (magnitude).}
	\label{Tag_Front_Little-Void}
\end{figure}

\subsection{Front corridor (tag 18)}
A corridor-like structure clearly visible in Figure \ref{Paper_Ingresso_Principale_3} (b) located just behind the external V-shaped structure depicted in Figure \ref{Paper_Ingresso_Principale_3} (a). The corridor is detected and its tomographic representation il depicted in Figure \ref{Paper_Ingresso_Principale_3} (b). 

\subsection{Big-void (tag 19)}
A large structure whose shape resembles a parallelepiped (number 19 of the 3D model). This object appears to be connected to structure 10 by means of a double horizontal connection (number 20 of the 3D model). The reference tomography images are shown in Figures \ref{IMG_2108} (b) \ref{IMG_2053} (b) and , while the 3D models are depicted in Figure \ref{IMG_2108} (a) and \ref{IMG_2053} (a). The large red target 1 visible in Figure \ref{IMG_2053} (b) is a false alarm, generated by the south-west ascending angle of the pyramid \cite{li2021stress}.
\begin{figure} 
	\centering
	\includegraphics[width=15.0cm,height=6.0cm]{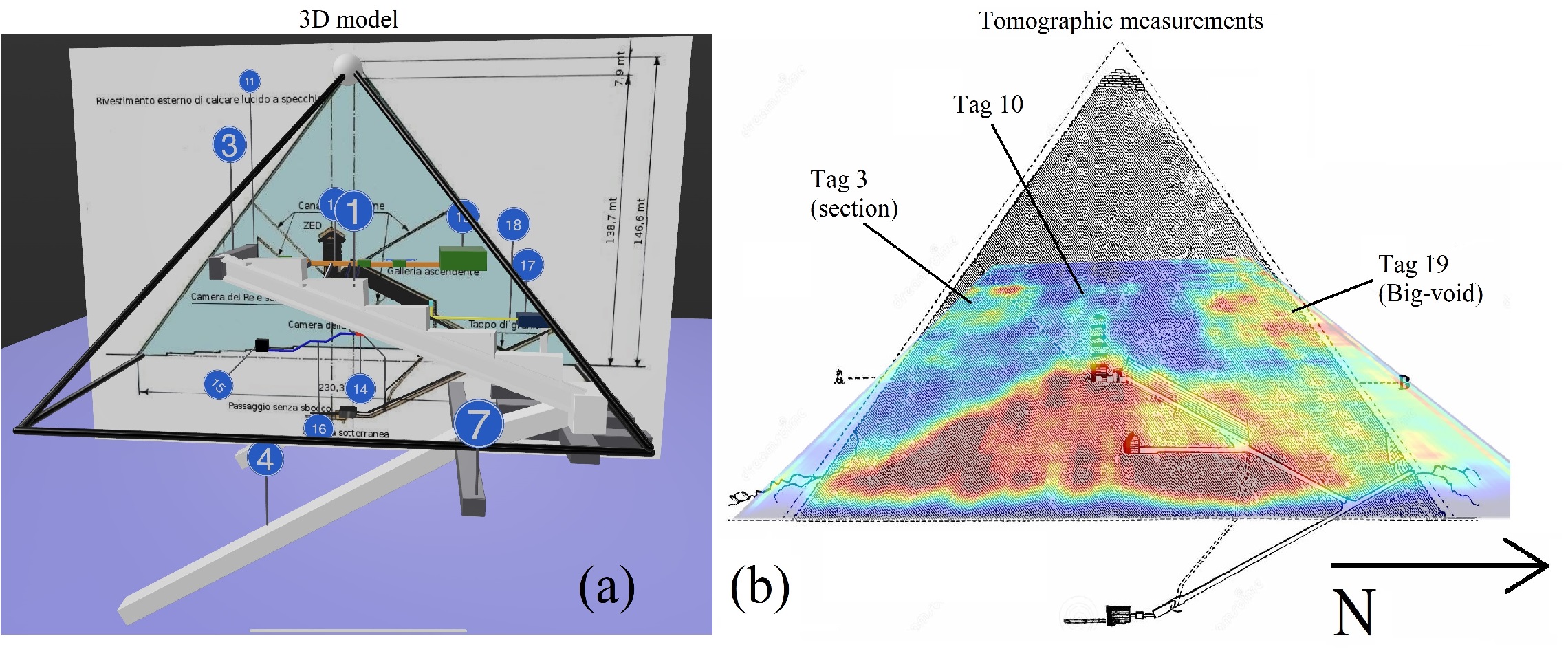}
	\caption{Tags association from tomography to 3D model. (a): 3D model of Khnum-Khufu. (b): Tomographic reconstruction (magnitude).}
	\label{IMG_2108}
\end{figure}
\begin{figure} 
	\centering
	\includegraphics[width=15.0cm,height=6.0cm]{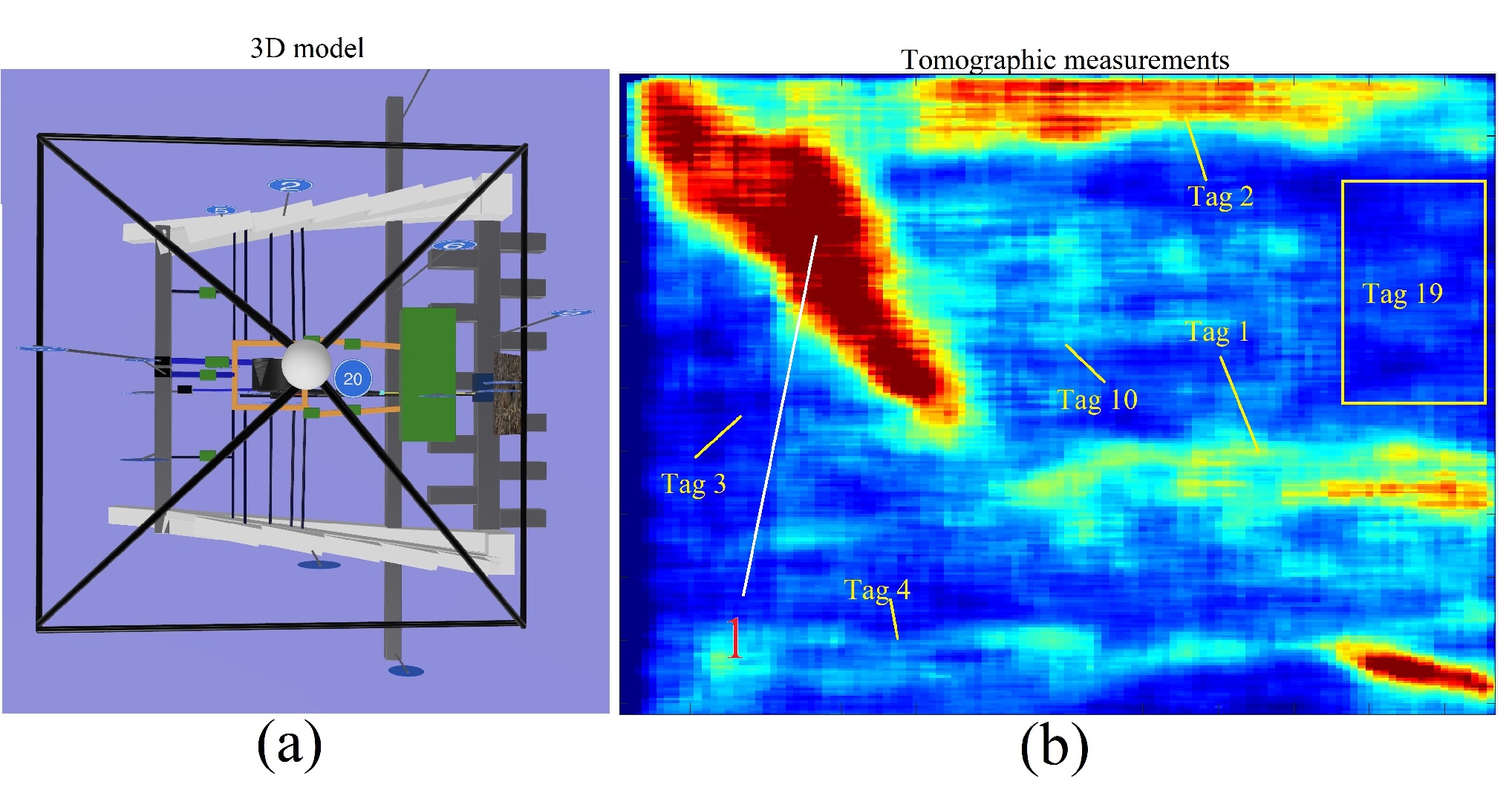}
	\caption{Tags association from tomography to 3D model. (a): 3D model of Khnum-Khufu. (b): Tomographic reconstruction (magnitude).}
	\label{IMG_2053}
\end{figure}

\subsection{ZED-Big-void double connection (tag 20)}
Structure 19 (big-void) is connected to the large corridor 3 to the south, at the height of the big-void, via two oblique corridors. The reference tomography is that shown in Figure \ref{Tags_Generic_4} (b), while the reference 3D model is that in Figure \ref{Tags_Generic_5} (a). 

\subsection{Metric determination}
The final objective of this study is to provide approximate measurements of the structures detected using the Doppler SAR tomography technique. The measurements that we propose are expressed in metres and are affected by an error that we have estimated to be very low, with respect to the actual measurement of the structures, according to the particular methodology we used. The dimensions are proposed in Figures \ref{Cad_con_Misure_1}, \ref{Cad_con_Misure_2} and \ref{Cad_con_Misure_3}. The measurements we suggest also include the thickness of the material used to construct them and are not to be intended as mere empty space. With regard to obtaining only empty space, research is currently underway in order to improve the technique and find a way to distinguish solid spaces from hollow spaces. The measures reported are evaluated by using Tape Measuring Wall Area software (http://www.pictureenginecompany.com/MeasureEngine/Promo.html) emploing as internal standard the pyramid's base lenght and are in accordance whith the results afforded by Sar data.

\begin{figure}[htp]
	\centering
	\includegraphics[width=12.0cm,height=10.0cm]{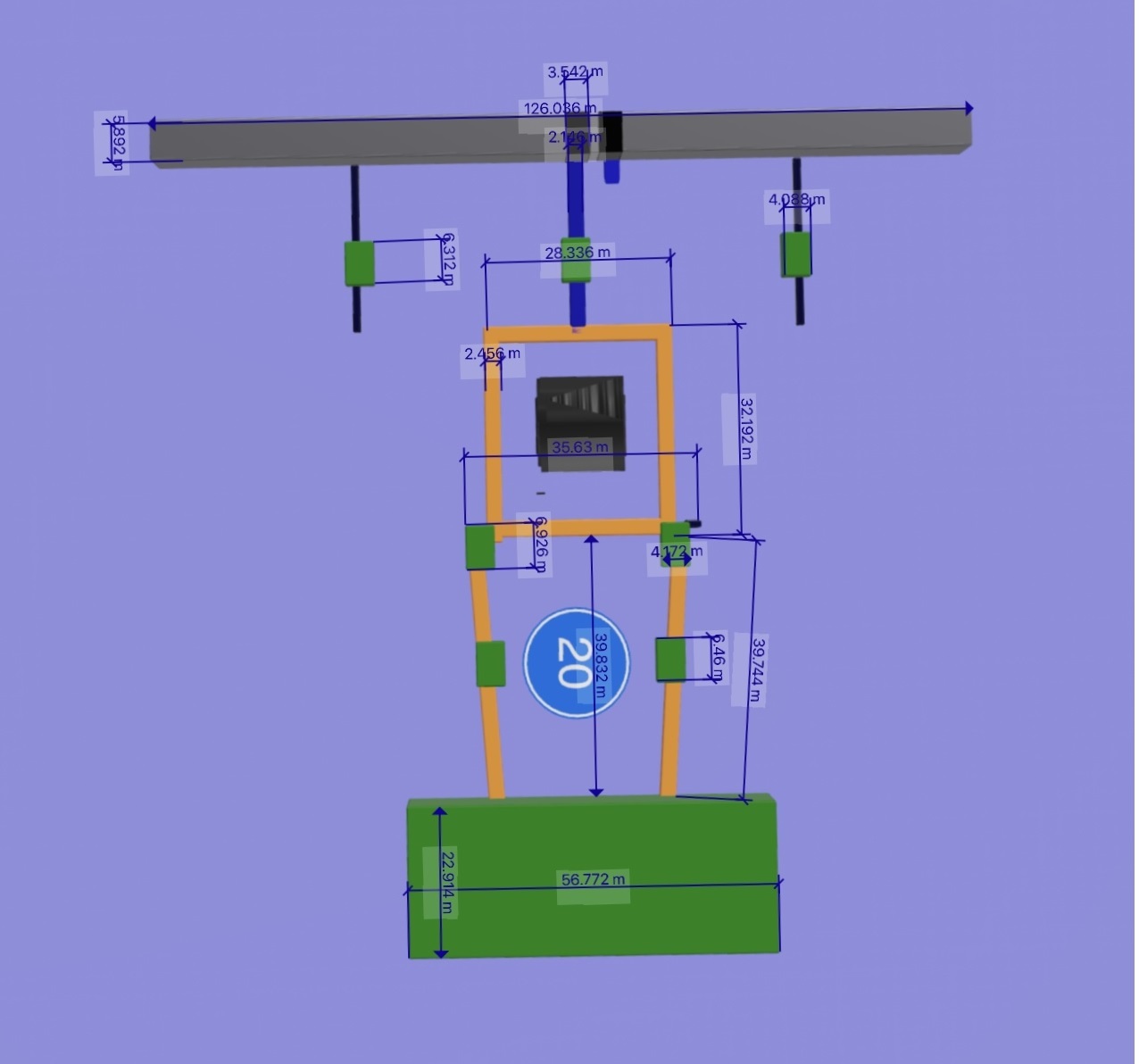}
	\caption{Measurements of the detected facilities of the pyramid. The numbers shown after the comma can not be significant.}
	\label{Cad_con_Misure_1}
\end{figure}

\section{Discussion}\label{Discussion}

\subsection{Data analysis}
Here we proceed to data analysis, while being aware that some aspects of the internal structure of the pyramid of Khnum-Khufu still need to be clarified, it is possible, in our opinion, to attribute a meaning to the internal structures of the monument, taking into account all the data relating to previous research in this field.
It is possible to underline how our 3D reconstruction, even if for now it does not claim to indicate the true measurements of the objects shown, (even if the SAR technique can accurately evaluate this parameter), is following some facts exposed in previous research.
Some researchers have shown how on the North edge of the East side of the pyramid at ground level, there is a thermal anomaly that suggests the presence of a room and a corridor located a few meters from the external wall of the monument \cite{ivashov2021proposed,marini2018real,bross2022tomographic,aly2022simulation}.
These data are in agreement with our analyses which predict, at that point, the presence of a room as a link between the two ramps 1 and 4 of our 3D model.
The microgravimetry data, carried out on the pyramid by different research groups \cite{ivashov2021proposed,yoshimura1987non,lheureux2010analyse,bui2011imaging}, can show us how, under the floor of the King's room, there is a lack of homogeneity possibly attributable to structures such as those hypothesized by us (see points 12 and 13 of the 3D model). The presence of rooms located under the King's room is also amply documented by many photographic finds on the Web.
The presence of ramps, placed inside the pyramid and which we highlight with certainty for the first time (numbers 1, 2, 3, and 4 of the 3D model), had already been postulated \cite{tasellari2013great} and partly detected by electrogravitic measurements carried out in 1998 \cite{bui1988application}.
The presence of further rooms located near the Queen's room had already been postulated in the past by some researchers \cite{bui1988application,blindenberg3d}, based on intuitions not confirmed, however, by objective evidence.
Authors of \cite{blindenberg3d} also postulated the presence of a horizontal passage placed between the original entrance of the pyramid and the Great Gallery, in place of structure number 17 which we highlight in this work; moreover, in this work, the presence of a room located immediately after the entrance to the great pyramid is highlighted. This void also seemed to be confirmed by muon spectroscopy \cite{houdin2007construction} carried out by researchers at the University of Nagoya in 2017 and indicated with the name of "Small Void".
It must be strongly emphasized that, while the analyses used up to now in the attempt to describe the objects inside the pyramid, gave only the possibility of making indirect hypotheses, the SAR methodology visibly produces direct evidence of the geometries inside the objects that can be analyzed. In contrast, the muon spectroscopy used in 2017 \cite{houdin2007construction} would not have been deemed reliable by the Egyptian Supreme Council of Antiquities, providing controversial data \cite{morishima2017discovery}.
No trace of the presence of a structure identified as a Big Void, put in evidence by muonic spectroscopy \cite{houdin2007construction}, was in our hands detected by SAR.
On the other hand, the SAR technique, allows us of making more observations using different starting geometries, thus being able to see the same structures inside the pyramid, from different points of view, and the only possibility of mistake lies in the visual interpretation of the data obtained but does not invalidate the presence or absence of specific structural elements. 
The geometries of the objects highlighted by the SAR in this paper, also make us reasonably exclude any errors due to the possible non-homogeneity of the materials employed during the construction of the pyramid interiors.

\begin{figure}[htp]
	\centering
	\includegraphics[width=15.0cm,height=12.0cm]{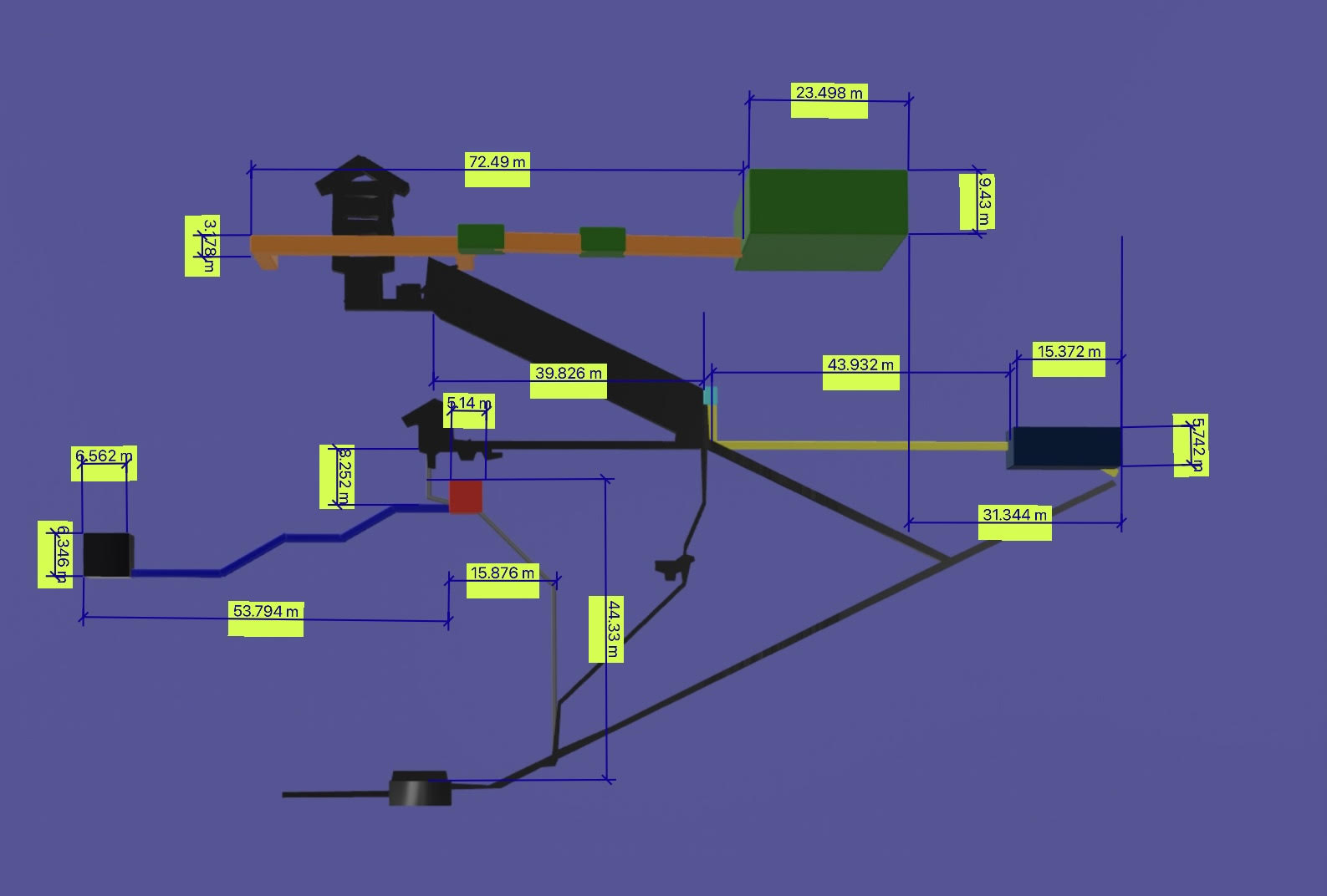}
	\caption{Measurements of the detected facilities of the pyramid. The numbers shown after the comma can not be significant.}
	\label{Cad_con_Misure_2}
\end{figure}

\subsection{Data interpretation}
On the basis of the above, it appears necessary to provide a plausible explanation as a key to understanding the function of the structures found inside the pyramid, taking into account that this interpretation is only intended to be a starting point for further interpretative ideas that could arise from a serene discussion at the level of the scientific community.
The authors' vision starts from what we have already previously published in \cite{WinNT} and widely discussed \cite{malanga1997cheope,whissell2007developmental,st2007enhanced,mulligan2012experimental,hassaan2016mechanical,henry2014monitoring,jouniaux2012electrokinetics,davidovits2009pharaohs,barsoum2006microstructural}.
Starting from the observation of the outside of the three pyramids of the Giza plateau, for the first time, we were able to establish that the three pyramids of Khnum-Khufu, Kefren and Menkaure have eight sides. This feature, known only for the larger pyramid, is now extended to the other two. According to the authors of \cite{morishima2017discovery}, the idea that the pyramids of the Giza plateau had this characteristic is due to the need to convey, in an orderly manner, the water that flowed along the faces of the pyramidal structures.
In the case of the pyramid of Khnum-Khufu, which we have analyzed in depth, it can be assumed, in analogy with other authors \cite{bhatt2019subcontinent,zheng2017vibration,jana2007great}, that it was surrounded by an enormous basin full of water, which allowed the circulation of some boats. These boats were used by some attendants with the task of bringing the water to about 90 meters high, pouring it into the South shaft by using many rotating stones probably similar to the Sabu diorite stone \cite{hassaan2016mechanical}. 
The SAR tecnique allows us to put in evidence that the shape of this monument do not resemble a perfect pyramidic form because the presence of a double changing in slope: the first of which of 14.5 ca. degrees at approximately 20 meters of high and the second one of 6  degrees ca. at approximately 100 meters of high.
The Nile river's water should have filled the basin up to the height of the first change of slope of the pyramid, thus allowing the Egyptian boats not to get stuck with the keel on the side of the pyramid itself
The water would have invaded the King's chamber but having reached the height of the granite basin inside the chamber (often referred to as the sarcophagus), it would not have exceeded that level in height and would have instead risen in the North shaft, whose entrance is placed at the same height as the basin, creating an air seal that effectively airlocked the room. The King's chamber in fact hermetically sealed would have caused excess water to rise up the North shaft.
The Queen's chamber would also be filled with water, up to the height of the shafts, by means of two connections to the shafts of the King's chamber, probably located in rooms 19 and 11, building a closed circuit, called Quincke's tube \cite{hixson1963quincke}.
As also proposed by other authors \cite{WinNT_2}, the pyramid, with its megalithic structure was placed in vibration by the wind and the low frequencies thus developed, which acted as a low pass filter allowing only low frequencies to bounce back on the roof of the Zed towards the King's chamber \cite{malanga1997cheope}.
Such a room would behave like an air-filled bottle of Helmholtz \cite{liu2021gradually}, in which the granite basin acted as a bottleneck. The walls of the basin, vibrating at low and precise frequencies, linked to the internal and external measurements of the basin itself, proportional to multiples of $\pi$ and the Golden Ratio $\phi$ \cite{malanga1997cheope}, would have caused the water contained in the Quincke's circuit to vibrate.
These frequencies, traveling through the closed circuit of Quincke's tube, at about 1400 m/sec, (speed of sound in the water), reached the Queen's chamber, where the height of the water could not exceed the height of shafts from the floor. A particular frequency could be developed, suitably amplified by the correct dimensions of the niche present in the West wall, which acted as a sound box for a musical instrument, releasing into the air a sound frequency that was able to interact with a cylindrical container, placed on the floor of the room, traces of which are still visible \cite{smyth1877our}.
This cylindrical container, probably made of wood, was put into resonance by the obtained low frequency.
Two individuals were placed both in the basin of the King's chamber and in the cylindrical container, in the Queen's chamber and appropriately treated with this low sound frequency for curative and religious purposes \cite{till2019sound}.
At the end of the procedure, the King's chamber was emptied by letting the water out of the Great Gallery and conveying it towards the room called "Grotto" towards the "Unfinished" chamber which brought the water back through a path in the floor, now occluded by debris, to the Nile.
Subsequently, the Queen's chamber was emptied in two steps: first a granite "plug" in the corridor leading to the room was removed: (this passage actually has a slight hydraulic slope towards the Great Gallery) and the water was made to flow out. At the floor of the Great Gallery, where it was conveyed towards the "Grotto". Subsequently, a plug placed in the floor was removed to finish the emptying of the room.
The water thus conveyed through the hole in the floor, highlighted in a book published in 1877 \cite{smyth1877our}, allowed the liquid to enter the room which, in our 3D reconstruction, corresponds to the number 14, eventually reaching the "Unfinished" room. and returning to the Nile.
The "Grotto" and room 14 are, in our opinion, necessary to stop the fall of water by slowing down its speed, with a mechanism similar to a common water jet pump used in laboratories to create vacuum in equipment, called Venturi's tube.
The evident traces of erosion due to water inside the pyramid rooms are in support of our interpretative hypothesis.
The three boulders that today are wedged at the beginning of the oblique corridor leading to the Great Gallery, would have been used as "plugs" to block the access of water to the exit of the pyramid or from the Queen's chamber by making them flow in different positions as needed.
The existence of a passage 18 seems to be related with a little open room, never described by anyone but well tracked  by numerous photographic evidences, that appears located at the top of the entrance of the Great Gallery, probably employed as security exit.
The entire system of the ramps highlighted by the SAR could be interpreted as a gigantic resonant structure, having the purpose of equalizing any differences in vibration between the North and South part of the pyramid, with the aim of making the square structure reach number 10, placed around the Zed, an equalized vibrational signal.
Similarly, the complex structure number 9 identified immediately below the plane on which the pyramid rests, has a shape similar to structures used to absorb the effects of mechanical vibrations that are transmitted through the ground \cite{d2017mechanical}.
\begin{figure}[htp]
	\centering
	\includegraphics[width=15.0cm,height=9.0cm]{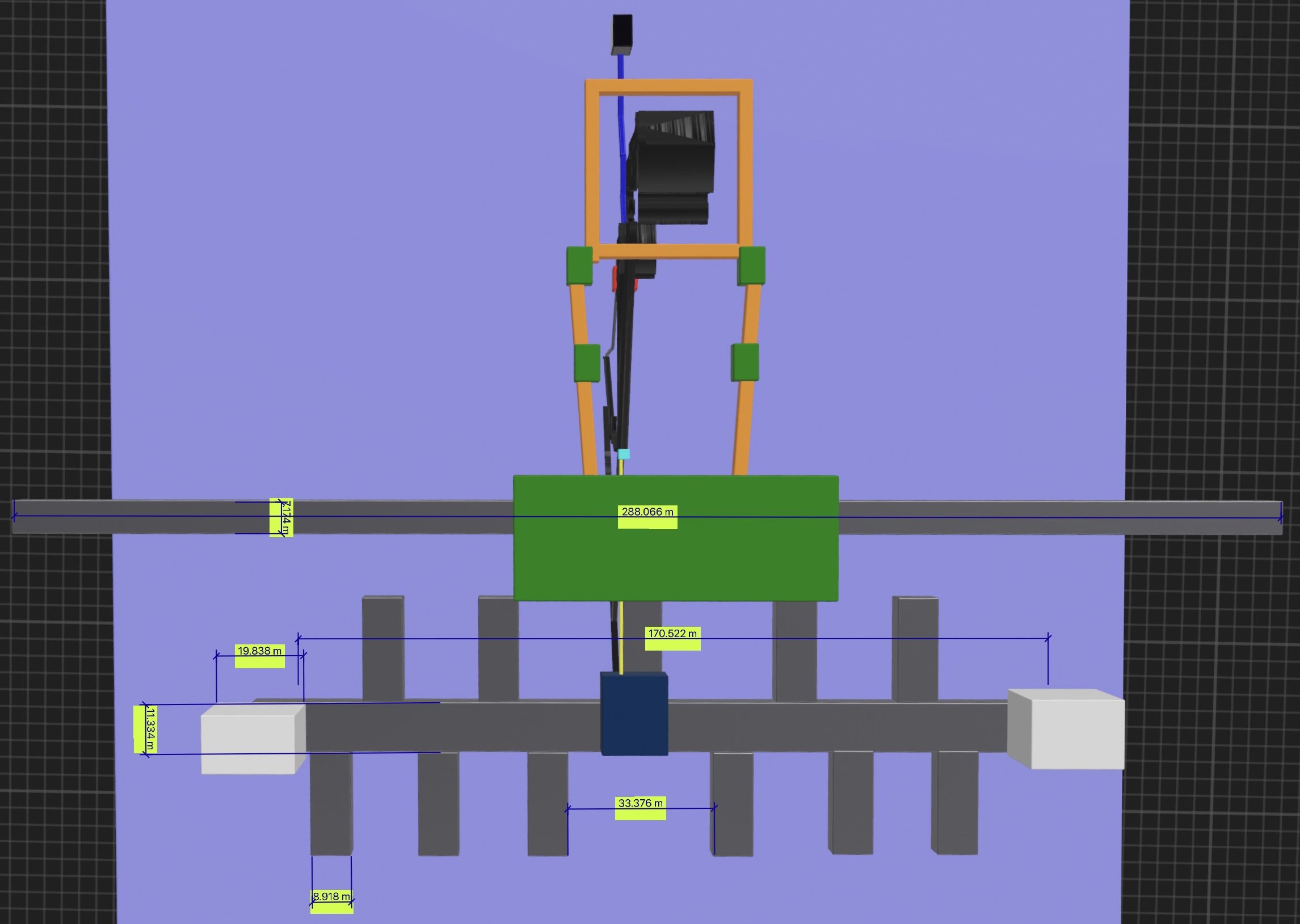}
	\caption{Measurements of the detected facilities of the pyramid. The numbers shown after the comma can not be significant.}
	\label{Cad_con_Misure_3}
\end{figure}

\begin{table}[tb!]
	\caption{List of the principal tomographic images.}
	\begin{center}
		\begin{tabular}{ |p{2cm}||p{4cm}|p{4cm}| }
			\hline
			Picture& Tomographic looking-direction & Tomographic line orientation \\
			\hline
			Figure \ref{Mask_6} & Eastern-side & Vertical \\
			Figure \ref{Mask_7} & Northern-side & Horizontal \\
			Figure \ref{Mask_7_1} & Western-side & Horizontal \\
			Figure \ref{SLC_2_1} & Eastern-side & Horizontal \\
			Figure \ref{Linea_Tomografica_3} & Western-side & Horizontal \\
			Figure \ref{Linea_Tomografica_2} & Northern-side & Horizontal \\
			Figure \ref{Linea_Tomografica_2_1} & Western-side & Vertical \\
			Figure \ref{SLC_3_1} & Eastern-side & Vertical investigation \\
			Figure \ref{Tomo_1_1} & Northern-Southern-side & Vertical \\
			Figure \ref{Tomo_1_2} & Southern-side & Horizontal \\
			\hline
		\end{tabular}
		\label{Tab_2}
	\end{center}
\end{table}

\begin{table}[tb!]
	\caption{Characteristics of the SAR acquisitions.}
	\begin{center}
		\begin{tabular}{ |p{2cm}||p{4cm}|p{7cm}| }
			\hline
			Structure number& Structure type & Structure name \\
			\hline
			1 & Corridor & Eastern ascending ramp \\
			2 & Corridor &  Western ascending ramp \\
			3 & Corridor &  Southern Corridor \\
			4 & Corridor &  Eastern descending ramp \\
			5 & Corridor &  Western descending ramp \\
			6 & Corridor &  Northern underground corridor\\
			7 & Corridor &  Northern-East underground corridor\\
			8 & Corridor &  Northern-West underground corridor\\
			9 & Complex structure & Northern underground complex-structure\\
			10 & Complex structure & ZED complex-structure \\
			11 & Room  & Eastern sarcophagus passage facility\\
			12 & Room  & Western sarcophagus passage facility\\
			13 & Room  & Bottom sarcophagus room facility\\
			14 & Room  & Queen's bottom room\\
			15 & Room  & Southern bottom room\\
			16 & Corridor & Southern Connection \\
			17 & Room  & Little-void \\
			18 & Corridor  & Front corridor \\
			19 & Room & Big-void\\
			20 & Complex structure & ZED-Big-void double connection\\
			\hline
		\end{tabular}
		\label{Tab_1}
	\end{center}
\end{table}

\section{Acknowledgments}\label{Acknowledgements}
We would like to thank Prof. Daniele Perissin for making the SARPROZ software available, through which many calculations were carried out more easily and quickly. We also thank the Italian Space Agency for providing the SAR data. We would also like to thank Dr. Riccardo Garzelli for conducting a deep revision of the English language and for having made crucial revisions to the entire article's structure and layout. The signal processing technique presented in this work has been submitted patent application in 04 July 2022, to Commerce Department of Malta, Industrial Property Registrations Directorate, patent application number 4451.

\section{Author contributions statement}
The authors contributed to all parts of this work

\section{Conclusions}\label{Conclusions}
In this paper we have shown how it is possible to use SAR micro-motion Doppler tomography in an advantageous, economical, non-invasive and rapid way to make a valid contribution in the study of the structure of ancient megalithic monuments such as the pyramid of Khnum-Khufu. We are aware that only by confirmation on the field of our findings can we validate our hypothesis. However it seemed logical to provide a hypothetical interpretation based on the data we collected that could serve as a starting point for future research. In the near future we would like to extend the SAR methodology to the investigation of the internal structure of other important monuments of the Giza plateau.}

\bibliographystyle{unsrt}  
\bibliography{references}

\end{document}